\documentclass[a4paper,12pt]{article}
\usepackage[14pt]{extsizes} 
\usepackage[mathscr]{eucal}
\usepackage{amsmath,amsfonts,amssymb,amsthm}
\usepackage{times}
\usepackage{hyperref}

\advance\textheight\topskip
\numberwithin{equation}{section} \makeatletter
\@addtoreset{equation}{section}

\newtheorem{thm}{Theorem}

 \newtheorem{prop}{Proposition}[section]
 
 \newtheorem{definition}[prop]{Definition}

\renewcommand{\tilde}{\widetilde}
\renewcommand{\hat}{\widehat}

\newcommand{\bref}[1]{\textbf{\ref{#1}}}

\newcommand{\p}[1]{|#1|}
\newcommand{\gh}[1]{\mathrm{gh}(#1)}

\newcommand{\dv}{\mathrm{d_v}}
\newcommand{\dd}{\partial}
\renewcommand{\d}{\partial}
\renewcommand{\dh}{\mathrm{d_h}}

\newcommand{\cF}{\mathcal{F}}

\renewcommand{\geq}{\,{\geqslant}\,}

\newcommand{\binner}[2]{%
  {\langle}\kern-4.15pt{\langle}#1{,}\,#2{\rangle}\kern-4.15pt{\rangle}}
\newcommand{\commut}[2]{[#1{,}\,#2]}

\newcommand{\half}{\mathchoice{%
    \ffrac{1}{2}}{\frac{1}{2}}{\frac{1}{2}}{\frac{1}{2}}}

\newcommand{\ffrac}[2]{\raisebox{.5pt}%
  {\footnotesize$\displaystyle\frac{#1}{#2}$}\kern1pt}

\newcommand{\dl}[1]{\mathchoice{\ffrac{\d}{\d #1}}{\frac{\d}{\d #1}}{\ffrac{\d}{\d #1}}{\ffrac{\d}{\d #1}}}

\newcommand{\st}[2]{{\overset{#1}{#2}}}

\newcommand{\vdl}[1]{\ffrac{{\delta}}{\delta #1}}

\newcommand{\manifold}[1]{\mathscr{#1}}
\newcommand{\manX}{\manifold{X}}

\newcommand{\CC}{\mathcal{C}}

\newcommand{\fR}{\mathbb{R}}

\newcommand{\fZ}{\mathbb{Z}}

 \def\cE{\mathcal{E}}
 \def\cG{\mathcal{G}}
 
 \def\cI{\mathcal{I}}

 \def\cL{\mathcal{L}}

\usepackage[left=3cm,right=2cm,
    top=3cm,bottom=3cm,bindingoffset=0cm]{geometry}

\newcommand{\tgamma}{q}
\newcommand{\hL}{\cL}
\usepackage{authblk}

\title{Presymplectic BV-AKSZ formulation of Conformal Gravity}
\author{Ivan Dneprov and Maxim Grigoriev}
\affil{Lebedev Physical Institute,\protect\\
  Leninsky Ave. 53, 119991 Moscow, Russia \vspace{1em}
  \\
  Institute for Theoretical and Mathematical Physics,\protect\\
  Lomonosov Moscow State University, 119991 Moscow, Russia}  \vspace{1em}

\date{}

\numberwithin{equation}{section}

\begin{document}

\maketitle

\begin{abstract}
We elaborate on the presymplectic BV-AKSZ approach to local gauge theories and apply it to conformal gravity.  More specifically, we identify a compatible presymplectic structure on the minimal model of the total BRST complex of this theory and show that together with the BRST differential it determines a full-scale BV formulation for a specific frame-like action which seems to be previously unknown. Remarkably, the underlying frame-like description requires no artificial off-shell constraints. Instead, the action becomes equivalent to the usual con\-formal gravity one, upon gauging away all the variables belonging to the kernel of the presymplectic structure. Finally, we show how the presymplectic BV-AKSZ approach extends to generic gauge theories.
\end{abstract}

\newpage

\tableofcontents

\section{Introduction}

It is well known that AKSZ sigma models~\cite{Alexandrov:1995kv} with finite dimensional target spaces are topological (= no local degrees of freedom). However, if one replaces the target space symplectic structure with a possibly degenerate presymplectic one, one arrives at an elegant AKSZ-like representation~\cite{Alkalaev:2013hta} for the familiar first-order Lagrangians and makes manifest the graded geometry underlying the frame-like formulations of various models, including gravity and higher spin fields.

It was later realised~\cite{Grigoriev:2016wmk} that the target space of a presymplectic AKSZ model arises as an equivalent reformulation of the total BRST complex of the local gauge theory in question, while the presymplectic structure can be seen as a BRST extension of the well-known on-shell presymplectic structure ~\cite{Kijowski:1979dj,Crnkovic:1986ex,Khavkine2012}, alias the presymplectic current. Moreover, the presymplectic current itself can be seen~\cite{Sharapov:2016sgx} as the descent completion of the Batalin-Vilkovisky (BV) symplectic structure.

Recently, it has become clear~\cite{Grigoriev:2020xec} that not only the Lagrangian, but also the entire BV formulation~\cite{Batalin:1981jr,Batalin:1983wj} of a given local gauge field theory, can be encoded in the presymplectic AKSZ formulation.\footnote{Mention also an alternative AKSZ approach~\cite{Canepa:2020rhu} to general relativity based on space+time decomposition.} In particular, the target space presymplectic structure defines a presymplectic structure on the space of AKSZ fields. Taking the symplectic quotient results in the BV symplectic structure which, together with the BV-BRST differential, determines the BV master action. This allows the presymplectic BV-AKSZ approach to be considered as a rather flex\-ible and general extension of the usual BV-BRST framework. In this approach a Lagrangian local gauge theory is described in terms of fields and gauge parameters that can be subject to differential constraints. For instance, the formulation of a Lagrangian system in terms of its stationary surface (equation manifold in the PDE theory terminology) is a typical example of such a description.

When applied to Lagrangian gauge theories extending gravity, this approach gives an interesting interpretation of the theories (and their underlying geo\-metry, e.g. Cartan geometry) in terms of the differential graded (pre)symplectic geo\-metry.  It is also worth mentioning the recent application of presymplectic AKSZ approach to constructing Lagrangian description of interacting higher spin theories~\cite{Sharapov:2021drr}. 

In this work we apply the full-scale presymplectic AKSZ approach to conformal gravity. The theory has been extensively studied~\cite{Kaku:1977pa,Fradkin:1985am,Maldacena:2011mk,Mannheim:2011ds,Rachwal:2022pfe, Edery:2006hg}, 
and has interesting properties, for instance it is renormalizable\footnote{Though the renormalization breaks the Weyl invariance and the theory is anyway non-unitary.} and  admits supersymmetric extension, see e.g.~\cite{Fradkin:1985am}. We concentrate on the minimal version (also known as minimal model) of the jet-bundle BV-BRST complex of the theory, which can be obtained via an additional reduction of the complex initially put forward in~\cite{Boulanger:2004eh,Boulanger:2007st,Boulanger:2007ab} (see also~\cite{Joung:2021bhf}), and propose a presymplectic structure therein which defines a full-scale BV-AKSZ description of the theory. This gives a new formulation of conformal gravity in terms of the connection of the conformal algebra, where the field content and gauge symmetries are fully determined by the underlying presymplectic $Q$-manifold and there is no need for artificial constraints like the familiar torsion-free constraint of the formulation~\cite{Kaku:1977pa}  (see also~\cite{Preitschopf:1998ei} for the discussion of frame-like formulations of conformal gravity).  In this respect our formulation can be considered natural and arising from first principles.

The graded geometry structures identified in this work give a new perspective on the underlying conformal geometry and more generally Cartan geometry. The formulation itself can be useful in the analysis of the associated QFT, as it naturally comprises the respective BV formulation and the  local BRST cohomology complex.

Finally, we present an extension of the presymplectic BV-AKSZ formulation to the case of not necessarily diffeomorphism-invariant systems, where the underlying $Q$-manifold is not a product of the AKSZ source and the target, and therefore one has to resort to more general objects known as gauge PDEs~\cite{Grigoriev:2019ojp} (see also~\cite{Barnich:2010sw,Grigoriev:2010ic} for earlier but less geometrical description). Gauge PDEs can be described as $Q$-bundles~\cite{Kotov:2007nr} over the shifted tangent bundle over the space-time. 

The paper is organized as follows: in Section~\bref{sec:AKSZ-1} we revisit the presymplectic BV-AKSZ formulation of diffeomorphism-invariant local gauge theories. In Section~\bref{sec:presymp-CGR} we briefly review conformal gravity and its well-known frame-like (gauge) formulation and propose a specific presymplectic structure on its BV-BRST complex, which yields the proper presymplectic BV-AKSZ representation of the system. Section~\bref{sec:gen-presymp} is devoted to a generalization of the approach to not necessarily diffeomorphism-invariant systems. Technical details are relagated to the Appendix. In addition, Appendix~\bref{sec:Kaku-str} contains an alternative proposal for the symplectic structure which is based on the well-known frame-like (gauge) formulation of conformal gravity.

\section{Presymplectic BV-AKSZ formulation}
\label{sec:AKSZ-1}
\subsection{AKSZ sigma models}
\label{sec:AKSZ-2}
The AKSZ approach~\cite{Alexandrov:1995kv} has been initially formulated for Lagrangian theories. Its extension to gauge systems at the level of equations of motion is straightforward and amounts to forgetting the symplectic structure and rephrasing everything in terms of the BRST differential rather than BV master action and it is instructive to start with this more general case. 

By AKSZ sigma model at the level of equations of motion we mean a pair of  $\fZ$-graded $Q$-manifolds, one of which, $(T[1]X,d_X)$, where $d_X$ is the de Rham differential of $X$ seen as a $Q$ structure on $T[1]X$,  is regarded as source (i.e. $X$ is a genuine space-time) and another one $(F,q)$
as the target. The $\fZ$-grading is denoted as $\gh$ in what follows. In particular, for  $(T[1]X,d_X)$ the ghost degree is just a form-degree. 

Fields of the model are maps from the source to the target.
It is convenient to introduce a trivial fiber bundle $(E,Q)=(T[1]X,d_X)\times (F,q)$ (here we mean product of $Q$-manifolds, i.e. $Q=q+d_X$ in the adapted coordinates and regard fields as section $T[1]X \to E$. Recall that such maps are by definition homomorphisms of the algebras of functions and, in contrast to supermaps,  they preserve the degree. The generalization to the case where such a bundle is trivial only locally (but still $Q=q+d_X$) is straightforward. More general situation is discussed in Section~\bref{sec:gen-presymp}.

If $\sigma:T[1]X \to E$ denotes a field configuration then the equations of motion read as
\begin{equation}
\label{aksz-eom}
    d_X \circ \sigma^*=\sigma^* \circ Q\,.
\end{equation}
The infinitesimal gauge variation of the configuration $\sigma$ is given by
\begin{equation}
\label{aksz-gs}
    \delta\sigma^*=d_X \circ \xi_\sigma^*+\xi_\sigma^* \circ Q\,
\end{equation}
where $\xi_\sigma^*:\CC^\infty(E)\to \CC^\infty(T[1]X)$ is a gauge parameter map, which has a degree $-1$ and satisfies
\begin{equation}
    \xi_\sigma^*(fg)=(\xi_\sigma^*(f))\sigma^*(g)+(-1)^{\p{f}}\sigma^*(f)\xi_\sigma^*(g)\,.
\end{equation}
In a similar way one defines gauge for gauge symmetries. The above is just a coordinate-free reformulation of the equations of motion and gauge symmetries encoded in the BRST differential of the AKSZ sigma models. Details can be found in~\cite{Grigoriev:2019ojp}.

A remarkable feature of the AKSZ construction is that 
the full-scale BV formulation (at the level of equations of motion) is also encoded in the AKSZ data.
More precisely, the space of fields, ghosts, and antifields is given by super-maps from $T[1]X$ to $F$ (super-sections of $E$) so that the space of sections is recovered as the degree $0$ component (the body) of the space of super-sections. The coordinates (fields and antifields) on the space of supermaps can be introduced as:
\begin{gather}
    \hat\sigma^*(\psi^A)=\st{0}{\psi}{}^A(x)+\st{1}{\psi}{}^A_{\mu}(x)\theta^\mu+\st{2}{\psi}{}^A_{\mu\nu}(x)\theta^\mu\theta^\nu+\ldots\,,    
\end{gather}
where $\hat\sigma^*$ is a pullback map induced by super-section $\hat\sigma$ and we made use of local coordinates $\psi^A$ on $M$ and the adapted local coordinates $x^\mu,\theta^\mu$ induced by local coordinates $x^\mu$ on $X$. Note that form-degree $k$ components $\st{k}{\psi}{}^A_{\mu_1\ldots \mu_k}$ carry ghost degree $\gh{\psi^A}-k$. The space of maps is recovered by setting to zero all the coordinates of nonvanishing degree.

In the field-theoretical terms the BRST differential of the AKSZ sigma model can be written as:
\begin{equation}
\label{AKSZ-diff}
    s=\int d^nx d^n \theta \big(q^A(\psi(x,\theta))-d_X\psi^A(x,\theta)\big)\vdl{\psi^A(x,\theta)}\,,
\end{equation}
where $n=\dim{X}$. 

If in addition to $Q$-structure the target space is equipped with the compatible symplectic structure $\omega$ of degree $\gh{\omega}=n-1$, i.e. $\omega$ is nondegenerate, $d\omega=0$ and $L_q\omega=0$ then the theory described by the AKSZ model is Lagrangian. More specifically, introducing $\hL$ via $i_q\omega+d\hL=0$ the action can be written as:
\begin{equation}
\label{AKSZ-action}
S[\sigma]=\int_{T[1]X} [(\sigma^*\, \chi)(d_X)+\sigma^*(\hL)]\,,
\end{equation}
where $\sigma$ is a section: $T[1]X \to E$ and we have introduced one form presymplectic potential $\chi$ such that $d\chi = \omega$. Here $(\sigma^*\, \chi)(d_X)$ denotes evaluation of a 1-form $\sigma^*\, \chi$ on the vector field $d_X$.

It turns out that just like the BRST differential the BV master action can also be immediately written just in terms of $q,\omega,d_X$. More precisely, it is given by 
\begin{equation}
\label{BV-like}
S_{BV}[\hat \sigma]=\int_{T[1]X} [(\hat \sigma^*\,\chi)(d_X)+\hat\sigma^*(\hL)]\,,
\end{equation}
i.e. has the same form as the classical action but with $\sigma$ replaced with $\hat\sigma$. It satisfies the master equation with respect to the odd Poisson bracket of degree $1$ induced by $\omega$ on the space of super-sections~\cite{Alexandrov:1995kv}.

If the target space is finite-dimensional and $n>1$ the AKSZ sigma model is topological.\footnote{For $n=1$ AKSZ sigma-models describe  constrained Hamiltonian systems with trivial Hamiltonians~\cite{Grigoriev:1999qz} , typical example of which are (spinning) relativistic particles. It is debatable whether it is natural to call such models topological.} However, generic diffeomorphism-invariant gauge theories can be represented as AKSZ sigma models at the price of considering infinite-dimensional targets. For instance taking as $M$ a fiber of the jet-bundle of BV-BRST formulation for a diff-invariant gauge system and as $q$ the restriction of its BV-BRST differential to the fiber (in the diff-invariant case the BV-BRST formulation can be locally encoded in the fiber, see e.g.~\cite{Barnich:1995db,Brandt:1996mh,Barnich:2010sw}) the respective AKSZ sigma model is equivalent to the initial gauge theory~\cite{Barnich:2010sw} (see also~\cite{Grigoriev:2010ic,Grigoriev:2012xg} for the Lagrangian version). Various equivalent (AKSZ) formulations can be arrived at by equivalent reductions of the target $Q$-manifold.  

\subsection{Presymplectic BV-AKSZ formulation}
\label{sec:presymp-AKSZ}

It turns out that relaxing the nondegeneracy condition for $\omega$ gives a natural generalization of AKSZ sigma models. By presymplectic AKSZ sigma model we mean the same data $(T[1]X,d_X)$ and $(F,q)$, where $F$ is equipped with not necessarily invertible 2-form $\omega$ satisfying $\gh{\omega}=n-1$, $L_q\omega=0$, $d\omega=0$.

An important difference with the usual AKSZ is that we do not require that the underlying gauge system is necessarily determined (at the level of equations of motion) by the BRST differential $s$ given by~\eqref{AKSZ-diff}. Instead, the gauge system is determined by the action functional~\eqref{AKSZ-action}, where some of the fields (on which the action doesn't really depend) are gauged away. It turns out that a variety of gauge theories can be  represented in this way~\cite{Alkalaev:2013hta}. Now we give a systematic exposition of BV-AKSZ formulation following~\cite{Grigoriev:2020xec}.

To make the definition of a presymplectic BV-AKSZ formulation more precise let us restrict for the moment to
a coordinate patch $U\subset X$ and use the adapted coordinates $x^\mu,\theta^\mu, \psi^A$. Then it is convenient to represent the space of supermaps $Smaps(T[1]U,F)$ as $Smaps(U,\bar F)$, where $\bar F=Smaps(\fR^{n}[1],F)$. A generic supermap from $\bar F$ is given by:
\begin{equation}
\label{scoord}
    \hat\sigma^*(\psi^A)=\sum_{l=0}^n \frac{1}{l!}
    \st{l}{\psi}{}^A_{\mu_1\ldots \mu_l} \theta^{\mu_1}\ldots \theta^{\mu_l} \,,
\end{equation}
so that as coordinates on $\bar F$ one can take $\st{l}{\psi}{}^A_{\mu_1\ldots \mu_l}$, $l=0,\ldots n$, $n=\dim{X}$. In what follows we use $\Psi^I$ as a collective notation for the coordinates of $\bar F$. Applying the same construction to all the coordinate charts of $X$ results in the locally-trivial fiber bundle $\bar E$ over $X$ with a typical fiber $\bar F$.~\footnote{Note that although $E$ is assumed trivial, $\bar E$ can be only locally-trivial if $TX$ is not globally-trivial.}

The 2-form $\omega$ on $F$ defines a $2$ form $\bar\omega$ on $\bar F$ in the standard way:
\begin{equation}
\label{bar-omega}
    \bar\omega=\int d^n\theta\, \omega_{AB}(\psi(\theta))\, d\psi^A(\theta)d\psi^B(\theta)\,,
\end{equation}
where $d\psi^A(\theta)$ can be seen as a generating function for the differentials $d\Psi^I$. Note that this form transforms as a density on $U$ under the change of $x^\mu$-coordinates. It defines a density on $X$ with values in the  2-forms on the fiber, which is defined globally.\footnote{Another possibility to work with an invariantly-defined object is to define an $n+2$ form $\omega^{\bar E}=(dx)^n \bar\omega$ on the total space of $\bar E$.}

By construction $\bar\omega$ is closed and $\gh{\bar\omega}=-1$. In addition we assume that $\bar\omega$ is regular. This may involve taking as $\bar F$ only those supermaps where the rank of $\bar\omega$ is constant and maximal (this is exactly what one naturally does~\cite{Grigoriev:2020xec} in the case of GR and, as we are going to see,
in the case of conformal gravity). A regular closed two form defines an involutive distribution of its zero vectors, i.e. vector fields satisfying $i_V\bar\omega=0$. It is the standard fact that
the symplectic quotient (the space of the integrable submanifolds, which exists, at least locally) is  naturally a symplectic manifold which we denote $\bar G$ in what follows. Moreover, it can be realised (again, at least locally) as a submanifold transversal to the kernel distribution in which case the symplectic structure coincides with the restriction of $\bar\omega$ to the submanifold.  In this way we have arrived at the space of fields equipped with the odd symplectic structure of ghost degree $-1$, just like in the case of BV formalism. 

To have a complete BV formulation, in addition to the symplectic structure $\bar \omega$ on the space of fields we need a master-action satisfying master equation. To analyse this it is convenient to consider a jet bundle $J^{\infty}(\bar E)$, where $\bar E \to X$ is the introduced above fiber bundle with the typical fiber $\bar F$. The BV-like symplectic structure is then introduced in the standard way as an $(n,2)$ form
\begin{equation}
\omega^{pBV}=(dx)^n \bar\omega_{IJ} \dv\Psi^I \dv\Psi^J\,,
\end{equation}
where we used standard local coordinates on 
$J^{\infty}(\bar E)$, induced by trivialisation $\bar E|_U=T[1]U\times \bar F$. Note that this symplectic structure is defined globally on $\bar E$. More invariant description of the underlying geometry and its generalization is presented in section~\bref{sec:gen-presymp}.

When written  in terms of $J^{\infty}(\bar E)$ the integrand (over $X$) of the BV-like action \eqref{BV-like} reads as
\begin{equation}
\begin{gathered}
L^{BV}=K+\bar \hL\,, \\ K= (dx)^n\int d^n\theta \chi_A(\psi(\theta)) \theta^\mu D_\mu \psi^A(\theta)\,,\qquad \bar \hL= (dx)^n \int d^n\theta \hL(\psi(\theta))\,,
\end{gathered}
\end{equation}
where $D_\mu$ denote total derivatives on $J^\infty(\bar E)$. It turns out that the restriction of $L^{BV}$ to the jet sub-bundle associated to the symplectic quotient $\bar G$ realized as a submanifold of $\bar F$ satisfies all the conditions for the BV master action. 

To see this let us first introduce a natural prolongation $\bar  q$ of the vector field $q$ from $F$ to $\bar F$:
\begin{equation}
\bar q (\psi^A(\theta))=(q\psi^A)(\theta).
\end{equation}
Note that $\bar q$ is a Hamiltonian vector field on $\bar F$ with Hamiltonian $\bar \hL$, i.e. $i_{\bar q} \bar\omega-d\bar \hL=0$.
Now we can introduce a vertical evolutionary vector field $s$:
\begin{equation}
s=D^F+\bar q^{pr} \,,   
\end{equation}
where $\bar q^{pr}$ denotes a natural prolongation of $\bar q$ to $\bar E$ defined by requiring  $\commut{\bar q^{pr}}{D_\mu}=0$ and $D^F$ is a vertical evolutionary vector field defined through
$D^F\psi^A(\theta)=-\theta^\mu D_\mu(\psi^A(\theta))$. Of course, $s$ is just a BRST differential \eqref{AKSZ-diff} written in terms of $J^\infty(\bar E)$.

We have the following relations \cite{Grigoriev:2020xec}: 
\begin{equation}
\label{iDomega}
    i_{D^F}\omega^{pBV}+\dv K=\dh(\cdot)\,, 
    \qquad  i_{D^F} i_{D^F} \omega^{pBV}=\dh (\cdot)\,,
\end{equation}
as well as
\begin{equation}
\qquad i_{D^F} i_{\bar q^{pr}} \omega^{pBV}=\dh(\cdot)\,.
\end{equation}
which can be directly checked and amount to:
\begin{equation}
\label{pme}
    i_s \omega^{pBV}+\dv L^{BV}=\dh (\cdot)\,, \qquad  i_{s} i_{s} \omega^{pBV}=\dh (\cdot)\,.
\end{equation}
The above relations remain intact when restricted to $J^\infty(\bar G\times X)$ understood as a jet subbundle of $\bar E$ (recall that we assumed that the symplectic quotient $\bar G$ is realised as a submanifold of $\bar F$), where by restriction of $s$ we mean its projection induced by the projection $\bar F \to \bar G$ (recall that there is a natural projection to the symplectic quotient). If by some abuse of notations $s$ denotes its projection to $J^\infty(\bar G\times X)$ the restricted relations read as:
\begin{equation}
    i_s \omega^{BV}+\dv L^{BV}=\dh (\cdot)\,, \qquad  i_{s} i_{s} \omega^{BV}=\dh (\cdot)\,,
\end{equation}
where $\omega^{BV}$ is the restriction of $\omega^{BV}$ to $J^\infty(\bar G\times X) \subset J^\infty(\bar E)$. Because $\bar G$ is symplectic the above relation imply the classical master equation satisfied by $L^{BV}$. 

To summarize, we have shown how  a presymplectic BV-AKSZ system gives rise to a genuine BV gauge system. The only subtle point is whether a resulting BV master action is a proper solution to the master equation. This has to be addressed separately. 

\subsection{Target space and the total BRST complex}
\label{sec:target}

A natural question is what is the origin of the target space $Q$-manifold. It turns out that it is directly related to  the total BRST complex of the system.

Suppose we start with a jet-bundle BV formulation. The underlying bundle is $\cE\to X$, i.e. coordinates on its fibres are fields, ghosts, ghosts-for ghosts, \ldots, as well as antifields (For the moment we do not assume the system to be Lagrangian and hence in general there is no natural pairing between fields and antifields). The jet-space BRST complex is given by local functions on $J^\infty(\cE)$ with the differential being the BRST differential $s$, i.e. a degree $1$  vertical evolutionary vector field on $J^\infty(\cE)$. 

If the system in question is diffeomorphism-invariant (i.e. space-time repar\-met\-rizations are among the gauge symmetries encoded in $s$), differential  $s$ does not depend on $x$ and hence its restriction to a fiber define a $Q$-structure (which we keep denoting by $s$) on a typical fiber $\cF$ of $J^\infty(\cE)$. Moreover, locally on $X$, $Q$-manifold $(\cF,s)$ is equivalent to the total BRST complex of the system, i.e. the complex of horizontal forms on $J^\infty(\cE)$ with the differential being $\tilde s=\dh+s$ and the degree being ghost degree plus form degree. 

It is known~\cite{Barnich:2010sw} that in diff-invariant case the AKSZ sigma model (at the level of equations of motion) with the target $(\cF,S)$ (or its equivalent reduction) and source $T[1]X$ gives an equivalent formulation of the gauge system in question. Furthermore, the compatible presymplectic structure $\omega$ on $F$ can be traced back ~\cite{Grigoriev:2020xec} (see also \cite{Sharapov:2016sgx}) to the initial BV symplectic structure extended to a cocycle of the total BRST differential $\tilde s$. More detailed and general exposition of the relation is given in Section~\bref{sec:gen-presymp}.

\section{Presymplectic AKSZ form of conformal gravity}

\label{sec:presymp-CGR}

\subsection{Conformal gravity}
Conformal gravity is a well-known theory of gravity with an additional gauge invariance compared to GR: local Weyl invariace. The action for such theory is 
given by
\begin{equation}
\label{metric-action}
    S[g] = \int d^4x \sqrt{-g} W_{\mu \nu \rho \lambda}W^{\mu \nu \rho \lambda}\,,
\end{equation}
where $W_{\mu \nu \rho \lambda}$ is the trace-free part of the Riemann curvature $R_{\mu \nu \rho \lambda}$. The indices are raised and lowered by the metric $g_{\mu \nu}$. Weyl invariance implies that the action only depends on the conformal class of the metric $[g_{\mu \nu}] = g_{\mu \nu}/ \sim$, where $g_{\mu \nu} \sim g'_{\mu \nu} = e^{2\phi}g_{\mu \nu}$.
In field theoretical terms, the infinitesimal gauge symmetries of the theory are given by
\begin{equation}
    \delta g_{\mu\nu}=\xi^\rho \dd_\rho g_{\mu\nu}+g_{\mu\rho}\dd_\nu \xi^\rho+g_{\rho\nu}\dd_\mu \xi^\rho +2\omega g_{\mu\nu}\,,
\end{equation}
where $\xi^\mu$ and $\omega$ are parameters of the diffeomorphisms and Weyl transformations respectively. 

The equations of motion determined by $S[g]$ read as:
\begin{equation}
B_{\mu \nu}=0\,, \qquad  \text{where}  \qquad B_{\mu \nu} \equiv P^{\rho \lambda}W_{\mu \rho \nu \lambda} + \nabla^\rho \nabla_\rho P_{\mu \nu} - \nabla^\rho \nabla_\mu P_{\nu \rho}
\end{equation}
and $P_{\mu \nu}$ is the Schouten tensor (for more details and precise definitions see Appendix~\bref{sec:techincal}). Tensor $B_{\mu\nu}$ is known as Bach tensor and it is Weyl invariant so that, as they should be, EOMs are diffeomorphism and Weyl invariant.

\subsection{Frame-like formulation}

Just like in the case of Einstein gravity where Poincar\'e algebra is the maximal symmetry of a vacuum solution and the theory itself can be reformulated in terms of the Poincar\'e algebra connection, the maximal symmetry of a solution to conformal gravity is the conformal algebra $o(d,2)$ and the theory also admits a  formulation in terms of the $o(4,2)$-connection. Although we don't really employ it in what follows, we now briefly review the formulation initially put forward in~\cite{Kaku:1977pa}. For further details and pedagogical presentation see e.g.~\cite{Trujillo:2013saa}.

The usual basis in $o(4,2)$ given by translations $P_a$, special conformal transformations $K_a$, Lorentz transformations $J_{ab}$, and dilation $D$, where the commutation relation read as:
\begin{equation} \label{conf_alg}
\begin{gathered}    
    \commut{J_{ab}}{J_{cd}}= \eta_{bc}J_{ad} + \eta_{ad}J_{bc} - \eta_{ac}J_{bd} - \eta_{bd}J_{ac} \,,\\
    [J_{ab}, P_c] = \eta_{bc}P_a - \eta_{ac}P_b, \qquad [J_{ab}, K_c] = \eta_{bc}K_a - \eta_{ac}K_b \,,\\
    [P_a, D] = P_a, \qquad [K_a, D] = -K_a\,, \\ 
    [K_a, P_b] = -2(\eta_{ab}D + J_{ab})\,.
\end{gathered}
\end{equation}
Here $\eta_{ab}$ denotes a constant Minkowski metric with almost positive signature.

In this basis the components of the connection 1-form $A$ and its curvature $\Omega$ are introduced as:
\begin{equation}
    A =  e^a P_a + \omega^{ab}J_{ab} + f^a K_a + \lambda D
\end{equation}
and 
\begin{equation}
    \Omega = T^a P_a +R^{ab} J_{ab} +  S^a K_a + \Lambda D\,.
\end{equation}
Here and in what follows we systematically omit the wedge product symbol. 

The action of the conformal gravity in terms of a connection of the conformal group can be written in the following form \cite{Kaku:1977pa}:
\begin{equation}
\label{Kaku-action}
   S[e,f, \omega, \lambda] = \int R_{ab}R_{cd}\epsilon^{abcd}\,,
\end{equation}
where 
\begin{equation}
    R_{ab} = \st{(0)}{R}_{ab} + (e_a f_b - e_b f_a)
\end{equation}
is the Lorenz sector part of the full conformal curvature,
\begin{equation}
    \stackrel{(0)}{R}_{ab} = d\omega_{ab} + \omega_{a}{ }^{c}\omega_{cb}
\end{equation}
is the curvature of the Lorentz-subalgebra component of the connection, and $\epsilon^{abcd}$ is an invariant totally\--antisymmetric tensor with $\epsilon^{0123}=1$.

Here we do not discuss global space-time geometry and hence consider Lagrangians modulo total derivatives. In particular the Gauss-Bonnet term  $\st{(0)}{R}_{ab}\stackrel{(0)}{R}_{cd}\epsilon^{abcd}$ can be omitted and the action can be rewritten as:
\begin{equation} \label{action}
\begin{gathered}
    S = 
    \int (2\stackrel{(0)}{R}_{ab}(e_c f_d - e_d f_c) + (e_a f_b - e_b f_a)(e_c f_d - e_d f_c)\epsilon^{abcd} = \\ = \int (4\stackrel{(0)}{R}_{ab}e_c f_d + 4e_a f_be_c f_d)\epsilon^{abcd} = \\ = 4\int(d\omega_{ab}e_c f_d + \omega_{a}{ }^{k}\omega_{kb}e_c f_d + e_a f_be_c f_d)\epsilon^{abcd} \,.
\end{gathered}
\end{equation}
It turns out~\cite{Kaku:1977pa} that action~\eqref{action}
is equivalent through elimination of the auxiliary fields to the standard conformal gravity action in the metric formulation~\eqref{metric-action}, provided one in addition imposes the torsion constraint
\begin{equation} \label{TorsFree}
    T_a \equiv de_a + \omega_a{ }^b e_b + e_a \lambda = 0.
\end{equation}
This constraint is in fact algebraic (allows to uniquely express $\omega_{ab}$ in terms of $e_a$ and $\lambda$) and hence the system remains Lagrangian. Now we briefly recall how this works. Technical details can be found in Appendix~\bref{sec:techincal}.

Let us consider first equations  for $f_d$:
\begin{equation} \label{fEOM}
    \epsilon^{abcd} R_{ab} e_c = 0.
\end{equation} 
The curvature 2-form can be expanded as $R_{ab} = R_{abcd}e^c e^d$. Then equation \eqref{fEOM} implies $R^c{}_{bcd}=0$ which in turn gives:
\begin{equation}
    f = -\frac{1}{6}\stackrel{(0)}{R}
\end{equation}
and
\begin{equation} \label{Schouten}
    f_{b,\mu} = -\frac{1}{2}(e^d{ }_{,\mu} \stackrel{(0)}{ R^c{ }_{bcd}} - \frac{1}{6}e_{b,\mu}\stackrel{(0)}{R})\,,
\end{equation}
where $f = f^a{ }_\mu e_a{ }^\mu$ and $\stackrel{(0)}{R}=\stackrel{(0)}{R}{}^a{ }_{bac}\eta^{bc}$ ). One concludes that on-shell $f_{a,\mu}$ is an auxiliary field which coincides with the Schouten tensor (expressed through the frame field $e^a{ }_\mu$) and can be eliminated.

Taking into account $T_a=0$, the Bianchi identity $d\Omega+\commut{A}{\Omega}=0$ in the sector of translations give:
\begin{equation}
    e_c R^c{ }_a + e_a \Lambda = 0
\end{equation}
which, in turn, implies (see Appendix~ \bref{sec:techincal} for details) that the curvature in the sector of dilations vanishes
\begin{equation}
    \Lambda = d\lambda + e_af^a = 0\,.
\end{equation}

Let us turn our attention to gauge transformations.  Consider the following transformations:
\begin{equation}
\label{gs-A}
    \delta A = d\alpha + [A,\alpha]\,,
\end{equation}
under which $\Omega$ transforms as:
\begin{equation}
    \delta \Omega = [\Omega, \alpha]\,.
\end{equation}
Note that in general these are not gauge symmetries of the action.
It turns out that for $\alpha=b^a K_a$, i.e.  for $\alpha$ non-vansihing only in the sector of  "special conformal transformations", this transformation is a gauge symmetry of the action thanks to $T^a=0$.  These gauge  transformations are algebraic in the sector of $\lambda$ (dilation part of the connection):
\begin{equation}
    \delta \lambda = - e^a b_a\,.
\end{equation}
It follows $\lambda$ is pure gauge and can always be set to zero by a gauge transformation. In this gauge the torsion free condition \eqref{TorsFree} implies  that $\omega$ is just the Levi-Civita connection.

To summarize: action~\eqref{Kaku-action} supplemented with the torsion constraint~\eqref{TorsFree} is equivalent to 
the Weyl action through the elimination of generalized auxiliary fields (auxiliary fields + algebraically pure gauge fields). The equivalent reduction boils down to the following steps: 
\begin{enumerate}
\item Elimination of the auxiliary field $f^a{ }_\mu$ 
\item Algebraically solving zero torsion condition for the Lorentz connection, $\omega = \omega(e,\lambda)$
\item Using gauge symmetry \eqref{gs-A} with $\alpha =b^a K_a$, which is algebraic in the sector of $\lambda_\mu$,  to set $\lambda_\mu=0$. In this way we arrived to action~\eqref{metric-action} modulo a topological term but expressed in terms of the frame field $e_\mu^a$ rather than metric.
\item Using gauge symmetry \eqref{gs-A} with $\alpha =k^a J_{ab}$ to eliminate all the components of the frame field but the metric,\footnote{This step is analogous to the elimination of the metric from the CGR BRST complex and strictly speaking might require  assumptions. See Appendix~\bref{MetrEl}.}  giving the initial~\eqref{metric-action} in terms of the metric.
\end{enumerate}


\subsection{Jet-bundle BRST complex for CGR}

As we recalled in Section~\bref{sec:target} the target space of the AKSZ formulation of a given
diffeomorphism-invariant system originates from a fiber of the jet-bundle BV-BRST formulation of the system.
Now we recall this formulation for CGR in the metric-like form and construct a particularly useful minimal model for the respective total BRST complex.

Let us restrict to local analysis and introduce the respective structures using local coordinates. The bundle $\cE$ is that of the metric $g_{\mu \nu}(x)$, ghosts $\xi^\mu(x)$ associated to the diffeomorphism parameters and ghost $\lambda(x)$, associated to the parameter of Weyl rescalings. The associated coordinates on the jet-bundle $J^\infty(\cE)$ read as:
\begin{equation}
\label{loc-jet}
    x^\mu\, \quad g_{\mu \nu|\alpha...}\,, \quad \xi^\mu{}_{|\alpha...}\,, \quad  \lambda_{| \alpha...}\,.
\end{equation}
In these coordinates total derivatives $D_\rho$ have the standard form, e.g. $D_\rho g_{\mu \nu}=g_{\mu \nu|\rho}$.

The gauge part $\gamma$ of the BRST differential is determined by 
\begin{equation}
    \begin{gathered}
        \gamma g_{\mu\nu} = \xi^\rho  g_{\mu \nu|{\rho}} + \xi^{\rho}{ }_{|\mu} g_{\rho \nu} +  \xi^{\rho}{ }_{|\nu} g_{\rho \mu} - 2\lambda g_{\mu \nu}\,, \\
        \gamma \xi^\mu = \xi^\rho \xi^{\mu}{ }_{|\rho}\,, \\
        \gamma \lambda = \xi^\rho \lambda_{|\rho}\,,
    \end{gathered}
\end{equation}  
together with the conditions $\commut{D_\mu}{\gamma}$
and $\gamma x^\mu=0$ (i.e. it is a vertical and evolutionary vector field).

The total BRST complex is given by the local horizontal forms on $J^\infty(\cE)$ (which, as before,  we identify with functions on $J^\infty(\cE)$ pulled back to $T[1]X$. In local coordinates these are simply functions of new variables $\theta^\mu$, i.e. coordinate on the fibers of $T[1]X$, and coordinates \eqref{loc-jet}. The total BRST differential reads as
\begin{equation}
\tilde \gamma =\gamma+\dh\,, \qquad \dh=\theta^\mu D_\mu
\end{equation}
Its cohomology in form degree $n+k$, $k\geq 0$ are known to be locally isomorphic to the cohomology of $\gamma$ in the space of local functionals. 

The above complex is the total BRST complex for off-shell CGR, i.e. CGR without equations of motion imposed. Its on-shell version is obtained by replacing $J^\infty(\cE)$ by its subbundle $\cG$ determined by the equations of motion $B_{\mu\nu}=0$ and their prolongations $D_{(\alpha)}B_{\mu\nu}=0$. A possible alternative is to introduce antifields and to work in terms of the complete BRST differential $s=\gamma+\delta$. 

Because CGR is diff invariant the total BRST complex is locally equivalent to that of local functions on the typical fiber $\cF$ of $\cG$ equipped with $\gamma$ restricted to the fiber. Moreover, the AKSZ sigma model with the source $T[1]X$ and target $(\cF,\gamma)$ is locally equivalent~\cite{Barnich:2010sw} to the CGR defined at the level of equations of motion.

\subsection{Minimal model of the BRST complex for CGR} 

We are interested in the presymplectic BV-AKSZ formulation that is natural in the sense that it involves a minimal number of fields (among all  equivalent formulation of this type). In particular, the resulting formulation can be regarded as a canonical one (of course there remains a freedom of field redefinition).
To this end it is natural to consider as a target space a minimal $Q$-manifold that is equivalent to $(\cF,\gamma)$ in the sense of natural equivalence of $Q$-manifolds. It is known~\cite{Barnich:2010sw} that AKSZ sigma models (at the level of equations of motion) based on equivalent target $Q$-manifolds are equivalent in the sense of elimination of generalized auxiliary fields. Moreover, minimal $Q$-manifold corresponds to a frame-like formulation of the underlying model. Minimal models for various local gauge theories were discussed from different perspectives in e.g. \cite{Barnich:1995ap,Brandt:1996mh,Barnich:2010sw,Grigoriev:2020lzu} but the idea and somewhat implicit version appeared already in~\cite{Stora:1983ct,Manes:1985df,Vasiliev:1988xc}.

Let us recall the notion of equivalence for $Q$-manifolds. Here we follow the approach developed in~\cite{Barnich:2004cr,Barnich:2010sw,Grigoriev:2019ojp}. First of all we need the following:
\begin{definition} \label{contr_man}
A contractible $Q$-manifold is a $Q$-manifold of the form $(T[1]V,d_V)$, where $V$ is a graded linear space understood as a graded manifold and $d_V$ is a canonical differential (De Rham) on $T[1]V$. 
\end{definition}
For our present purposes it is enough to consider a rather restrictive version of equivalence: a $Q$-manifold $(N,q)$ is called an equivalent reduction of $(M,Q)$ if 
$(M,Q)=(N,q)\times (T[1]V,d_V)$, i.e. $M$ 
is a product $N\times T[1]V$ and $Q$ is a product $Q$-structure. This generates the equivalence relation for $Q$-manifolds. 

It follows from the definition of the equivalent reduction that there exists independent functions $w^a,v^a$ on $M$ such that $Q w^a=v^a$ and $(N,q)$ can be identified as a $Q$-submanifold determined by $w^a=0$ and $v^a=0$.
Indeed, as $w^a,v^a$ one can take linear coordinates on $T[1]V$ pulled back to $M$ by the canonical projection $M\to T[1]V$. Moreover, one can introduce a local coordinate system $\phi^i,w^a,v^a$ on $M$ where 
\begin{equation}
    Q \phi^i=q^i(\phi)\,, \qquad 
    Q w^a=v^a\,.
\end{equation}
In this form it is clear why the equivalent reduction from $(M,Q)$ to $(N,q)$ is often refereed to as  elimination of contractible pairs (in this form it has been extensively used in the context of local BRST cohomology for a long time, see e.g.~\cite{Henneaux:1990ua,Barnich:1995db,Brandt:1996mh,Barnich:2000zw,Brandt:2001tg}).

A particular equivalent reduction of the $Q$-manifold $(\cF,\gamma)$ in the case of off-shell CGR has been obtained by N. Boulanger in~\cite{Boulanger:2004eh,Boulanger:2007st}. More precisely, it was shown that all the derivatives of the metric except for Weyl curvature and its independent covariant derivatives can be taken as $w^a$  variables while the respective $v^a$ variables parameterize all the derivatives of ghosts of order 2 and higher.  As coordinates $\phi^i$ one takes: $\xi^\mu,\xi_\mu{}^\nu$ related to $\xi^\mu$ ghost and its 1-st derivatives, $\lambda,\kappa_a$
related to $\lambda$ ghost and its 1-st derivatives, and  $W_{\mu\nu\rho\sigma|\alpha\ldots}$ parameterizing Weyl tensor and its independent covariant derivatives. The action of $\gamma$ on the ghost degree $1$ variables is given by
\begin{equation}
\label{gamma-bul}
\begin{gathered}
    {\gamma} \xi_\mu = \xi^\rho {\xi}_{\rho \mu}\,, \\
    {\gamma} {\xi}_{\mu}{ }^\nu = {\xi}_{\mu}{ }^\rho {\xi}_{\rho}{ }^\nu + P^{ \lambda \nu}_{\mu \rho} \kappa_\lambda \xi^\rho + \frac{1}{2}\xi^\rho \xi^\lambda W_\mu{ }^\nu{ }_{\rho \lambda}\,, \\
    {\gamma} \kappa_\nu = {\xi}_\nu{ }^{\rho}\kappa_\rho  + \frac{1}{2}\xi^\rho \xi^\lambda C_{\nu \rho \lambda}\,, \\
    {\gamma}\lambda = \kappa_\rho \xi^\rho \,,
\end{gathered}
\end{equation}
where $P^{\lambda \nu}_{\mu \rho} = -g^{\lambda \nu}g_{\mu \rho} + \delta^\lambda_\mu \delta^\nu_\rho + \delta^\lambda_\rho \delta^\nu_\mu$,
and  denotes the Cotton tensor which can be expressed in our coordinates as $C_{\mu \nu \rho}=W^\alpha{}_{\mu\nu\rho|\alpha}$. In what follows we will also denote $C_{\mu \nu \rho | \alpha_1...} :=W^\alpha{}_{\mu\nu\rho|\alpha \alpha_1...}$ .  The action of $\gamma$ on the metric components reads as
\begin{equation}
    {\gamma} g_{\mu \nu} = 
    -2\lambda g_{\mu \nu} + {\xi}_\mu{ }^\rho g_{\rho \nu} + {\xi}_\nu{ }^\rho g_{a\rho}
\end{equation}
and on Weyl tensor and its covariant derivatives $\gamma$ acts in the following way:
\begin{equation}
\label{W-action}
    {\gamma} W= \xi^\mu R_\mu W + {\xi}^\mu{ }_\nu \Delta^\nu{ }_\mu W +  \kappa_\mu \Gamma^\mu W\,,
\end{equation}  
where
\begin{equation}
\begin{aligned}
        R_\beta W^{\mu}{ }_{\nu \rho \sigma| \alpha_1...\alpha_k} &=W^{\mu}{ }_{\nu \rho \sigma| \alpha_1...\alpha_k \beta}\,, \\
        \Delta^\beta{ }_\gamma W^{\mu}{ }_{\nu \rho \sigma| \alpha_1...\alpha_k} &= - \delta^\mu_\gamma  W^{\beta}{ }_{\nu \rho \sigma| \alpha_1...\alpha_k} + \delta^\beta_\nu  W^{\mu}{ }_{\gamma \rho \sigma| \alpha_1...\alpha_k}+\ldots \,. \\
\end{aligned}
\end{equation}
The action of $\Gamma^\mu$ can be determined from the  nilpotency of ${\gamma}$ and can be found in~\cite{Boulanger:2004eh}. For example, ${\gamma}^2\kappa_a = 0$ implies  the relation $\Gamma^\kappa C_{\mu \nu \rho} = W^\kappa{ }_{\mu \nu \rho}$ which we need in what follows. We refer to~\cite{Boulanger:2004eh} for further details.

For our purposes we need to reduce the above $Q$-manifold even further. It turns out that under some assumption on allowed class of metric configurations one can eliminate the metric together with the symmetric component of $\xi_{\mu\nu}$. Indeed,
\begin{equation}
    {\gamma} g_{\mu \nu} = -2\lambda g_{\mu \nu} + {\xi}_\mu{ }^\rho g_{\rho \nu} + {\xi}_\nu{ }^\rho g_{a\rho} = -2\lambda g_{\mu \nu} +2 {\xi}_{(\mu \nu)}\,.
\end{equation}
We relegate the discussion of the conditions under which this gives an equivalent reduction (for instance this the case if the metric is Euclidean) to Appendix~\bref{MetrEl} and simply assume that these are fulfilled. 

In a suitable coordinate system $x^a$ we decompose  $g_{ab}=\eta_{ab}+h_{ab}$, with $\eta_{ab}$ being standard Minkowski metric and find
\begin{equation}
    \gamma h_{ab}=-2\lambda g_{ab}+2\xi_{(ab)}\,,\qquad \xi_{(ab)}=\half(\xi_a{}^c g_{cb}+\xi_b{}^c g_{ac})\,.
\end{equation}
So that $h_{ab}$ can be taken as $w^a$ variables in the equivalent reduction. The reduced $Q$-manifold is obtained by setting $h_{ab}=\gamma h_{ab}=0$, giving
\begin{equation}
    g_{ab}=\eta_{ab}\,, \qquad \xi_{(ab)}=\lambda \eta_{ab}\,.
\end{equation}
In what follows we use the antysymmetric matrix $\rho_{ab}=\half(\xi_{ab}-\xi_{ba})$ to parametrize the remaining components of $\xi_{ab}$.

To summarize, the reduced $Q$-manifold $(F,q)$ is coordinatized by
degree $1$ coordinates $\xi^a,\rho^{ab}, \kappa_b, \lambda$ and degree $0$ coordinates $W_{a(b}{}^c{}_{d|j \ldots )}$ (though  it is very convenient to work with the overcomplete set of $W$-coordinates $W_{abcd|j \ldots }$ as we keep doing). The action of the reduced $\gamma$ (denoted by $q$ below) on the ghost degree $1$ coordinates can be easily obtained from~\eqref{gamma-bul} and is given explicitly by \footnote{By setting  to zero $W$ and $C$ one arrives at the Chevalley-Eilenberg differential for conformal algebra, however in a slightly rescaled basis comparing to \eqref{conf_alg}}: 
\begin{equation} \label{gamma_min}
\begin{gathered}
    \tgamma \xi^a = \rho^{a}{ }_c\xi^c + \xi^a \lambda\,, \\
    \tgamma \rho^{a}{ }_b = \rho^{a}{ }_c \rho^{c}{ }_b + (\xi^a\kappa_b - \xi_b\kappa^a) + \frac{1}{2}\xi^c \xi^d W^a{ }_{bcd}\,, \\
    \tgamma \kappa_b = \kappa_c\rho^{c}{ }_b + \lambda \kappa_b + \frac{1}{2}\xi^c \xi^d C_{bcd}\,, \\
    \tgamma\lambda = \kappa_c\xi^c \,.
\end{gathered}
\end{equation}
The action of $\tgamma$ on $W$-coordinates is also easily inferred from~\eqref{W-action}. In particular, for the Weyl and Cotton tensors we explicitly get:
\begin{equation}
\label{WC-rel}
\begin{aligned}
        \tgamma W^a{ }_{bcd} =&~ \xi^k  W^a{ }_{bcd|k} - \rho_k{ }^a W^k{ }_{bcd}+ \\&~\rho_b{ }^k W^a{ }_{kcd} + \rho_c{ }^k W^a{ }_{bkd} + \rho_d{ }^k W^a{ }_{bck} + 2\lambda W^a{ }_{bcd}\,,\\
    \tgamma C_{abc} =&~  \xi^k  C_{abc|k} + \rho_a{  }^k C_{kbc}+ 
    \\&~\rho_b{  }^k C_{akc} + \rho_c{  }^k C_{abk} + 3\lambda C_{abk}  + \kappa_k W^k{ }_{abc} \,.
\end{aligned}
\end{equation}
It should be stressed that the second equation is not written in independent coordinates which are the properly symmetrized Weyl "derivatives". 

Furthermore, we need to switch to the on-shell version of this $Q$-manifold, where the Bach equation and its differential consequences are taken into account. The elimination of contractible pairs of $\gamma$ is not affected by this restriction as the equations do not affect ghosts and gauge non-invariant quantities.  In particular, all the variables entering \eqref{gamma_min} and \eqref{WC-rel} 
remain independent upon restriction to the surface of equations except for $C_{abc|k}$ whose traces vanish on the surface. Using the restriction of these coordinates to the surface as the coordinates therein one concludes that the above relations stay true. In this way we have arrived at the $Q$-manifold $(F,q)$ which is going to serve as a target space of the presymplectic AKSZ formulation of CGR.

\subsection{Presymplectic structure} \label{sec:presymp_struc}

As was discussed above the minimal $Q$-manifold $(F,q)$ of CGR should admit a compatible presymplectic structure of ghost degree $n-1$, which originates from the descent completion of the BV symplectic structure. More precisely,  the minimal $Q$-manifold can be seen as a $Q$-submanifold of total BRST complex of CGR and the presymplectic structure can be obtained as the restriction of this descent completion to the submanifold.\footnote{In this setup, this is just an explicit realization of the homotopy transfer.}

Instead of computing the descent completion and its transfer to the minimal $Q$-manifold $(F,q)$, which can be quite involved technically, we try to find such a structure by considering a generic ansatz of the closed 2-form of ghost degree $3$ and not involving higher Weyl tensors. By some trial and error one finds that the correct presymplectic structure  should involve Weyl tensor and Cotton tensors and the remaining ambiguity is fixed by $\tgamma$-invariance. The end result reads explicitly as:
\begin{equation}
\label{presymp-main}
\begin{gathered}
    \omega = \omega_W-2\omega_C\,, \\ \omega_W=d(\rho_{ab})d(W^{abnm}\epsilon_{nmpk}\xi^p\xi^k)\,,\qquad \omega_C=d(\xi_a)d(C^a{ }_{bc}\epsilon^{bcpk}\xi_p \xi_k)\,. 
\end{gathered}
\end{equation}
\begin{prop}
Presymplectic structure~\eqref{presymp-main} on $F$ satisfies: $d\omega=0$,  $L_\tgamma \omega=0$, and $\gh{\omega}=n-1$ and hence together with the homological vector field $\tgamma$ it defines a presymplectic AKSZ model.
\end{prop}
The first and the third properties are obvious. Let us give here main points of the proof of $L_\tgamma \omega=0$ (details are relegated to the Appendix~\bref{sec:presymp-details}). 
It is enough to show that $i_\tgamma \omega=d\hL$ for some $\hL$. We have the following relations:
\begin{multline} \label{iqomw}
    i_{\tgamma}\omega_W = d(\rho_{al}\rho^l{ }_b W^{abnm}\epsilon_{nmpk}\xi^p\xi^k) + (\xi_a\kappa_b - \xi_b\kappa_a)d(W^{abnm}\epsilon_{nmpk}\xi^p\xi^k) + \\ + \frac{1}{2}W_{abij}\xi^i\xi^j d(W^{abnm}\epsilon_{nmpk}\xi^p\xi^k) + d(\rho_{ab})\xi^j  W^{abnm}{ }_{|j}\epsilon_{nmpk}\xi^p\xi^k \,.
\end{multline}
It is nearly $dH_W$ with $H_W$ given by
\begin{multline}
    H_W = \rho_{al}\rho^l{ }_b W^{abnm}\epsilon_{nmpk}\xi^p\xi^k +   2\xi_a\kappa_b W^{abnm}\epsilon_{nmpk}\xi^p\xi^k + \\+W_{abij}\xi^i\xi^j W^{abnm}\epsilon_{nmpk}\xi^p\xi^k\,,
\end{multline}
with the  discrepancy being
\begin{multline} \label{Wdisc}
    i_{\tgamma}\omega_W - dH_W = {-2d(\xi_a)\kappa_b W^{abnm}\epsilon_{nmpk}\xi^p\xi^k} + \\+ {2\xi_a d(\kappa_b) W^{abnm}\epsilon_{nmpk}\xi^p\xi^k} + {d(\rho_{ab})\xi^j  W^{abnm}{ }_{|j}\epsilon_{nmpk}\xi^p\xi^k}\,.
\end{multline}
As for the second term $\omega_C$ we find:
\begin{equation} \label{iqomc}
\begin{gathered}
    i_{\tgamma} \omega_C = \rho_{an}\xi^n d(C^a{ }_{bc}\epsilon^{bcpk}\xi_p \xi_k) +  d(\xi^a)\rho_a{  }^n C_{nbc} \epsilon^{bcpk}\xi_p \xi_k + \\ + \xi_a \lambda d(C^a{ }_{bc}\epsilon^{bcpk}\xi_p \xi_k) + d(\xi_a) \lambda C^a{ }_{bc}\epsilon^{bcpk}\xi_p \xi_k + \\ + d(\xi_a)\xi^d  C_{abc|d}\epsilon^{bcpk}\xi_p \xi_k + d(\xi^a) \kappa_k W^k{ }_{abc} \epsilon^{bcpk}\xi_p \xi_k\,.
\end{gathered}
\end{equation}
Introducing $H_C$ via
\begin{equation}
    H_{C} = \rho_{an}\xi^n C^a{ }_{bc}\epsilon^{bcpk}\xi_p \xi_k + \xi_a \lambda C^a{ }_{bc}\epsilon^{bcpk}\xi_p \xi_k
\end{equation}
we find
\begin{equation} \label{Cdisc}
\begin{gathered}
    i_{\tgamma} \omega_C - dH_C = {-d(\rho_{an})\xi^n C^a{ }_{bc}\epsilon^{bcpk}\xi_p \xi_k} +{ \xi_a d(\lambda) C^a{ }_{bc}\epsilon^{bcpk}\xi_p \xi_k} + \\ + {d(\xi_a)\xi^d  C_{abc|d}\epsilon^{bcpk}\xi_p \xi_k} {- d(\xi^a) \kappa^k W_{akbc} \epsilon^{bcpk}\xi_p \xi_k}\,,
\end{gathered}
\end{equation}
so that 
\begin{equation} \label{finres}
    i_{\tgamma}(\omega_W - 2\omega_C)
    = d \hL\,, 
\end{equation}
where the "Hamiltonian" is given by:
\begin{multline}
\hL=H_W-2H_C=\\= \rho_{al}\rho^l{ }_b \hL=W^{abnm}\epsilon_{nmpk}\xi^p\xi^k+ 2\xi_a\kappa_b W^{abnm}\epsilon_{nmpk}\xi^p\xi^k -2\rho_{an}\xi^n C^a{ }_{bc}\epsilon^{bcpk}\xi_p \xi_k -\\-2\xi_a \lambda C^a{ }_{bc}\epsilon^{bcpk}\xi_p \xi_k +W_{abij}\xi^i\xi^j W^{abnm}\epsilon_{nmpk}\xi^p\xi^k\,.    
\end{multline}
Note that we have changed the sign convention for $\cL$ with respect to Section~\bref{sec:AKSZ-1} to comply with the standard conventions for curvatures. This is equivalent to changing the sign of $q$ from the very beginning and hence does not affect the BV master equation.

We are now ready to spell out an explict form of the action. Introducing component fields parameterizing $\sigma : T[1]X \to F$ via
\begin{equation}
\label{fields}
\begin{gathered}
\sigma^*(\xi^a)=e^a_\mu(x)\theta^\mu\,, \quad
\sigma^*(\rho^{ab})=\omega^{ab}_\mu(x)\theta^\mu\,,
\\
\sigma^*(\kappa^a)=f^a_\mu(x)\theta^\mu\,, \quad
\sigma^*(\lambda)= v_\mu(x)\theta^\mu\,, \\
\sigma^*(W_{abcd})=W_{abcd}(x)\,, \quad
\sigma^*(C_{abc})=C_{abc}(x)\,, \quad
\end{gathered}
\end{equation}
where by some abuse of notations we sometimes used the same letters to denote some target space coordinates and the respective fields. It can be convenient to treat $e,\omega,W,C$ as differential forms on $X$. With this identification the action~\eqref{AKSZ-action} can be written as:
\begin{multline} \label{Act}
    S[e,\omega,W,C] = \int_X  \big[(d\omega_{ab}+ \omega_{ac}\omega^c{ }_b+ e_a f_b - e_b f_a)W^{abnm}\epsilon_{nmpk}e^p e^k +\\+ W_{abcd}e^c e^d W^{abnm}\epsilon_{nmpk}e^p e^k
    -2 (de_a + \omega_{ad}e^d + e_a v)C^a{ }_{bc}\epsilon^{bcpk}e_p e_k\big]\,,
\end{multline}
where we omit the wedge product symbol. Note that before the symplectic reduction has been done the action is to be considered as that depending on all the degree zero fields. However here, we only spelled explicitly those fields on which it depends nontrivially. In particular, thanks to the symmetry properties of $W,C$, fields $v,f$ do not contribute (as we are going to see this is in agreement with the fact that $\dl{\lambda}$ and $\dl{\kappa}$ are in the kernel of the presymplectic structure). 

To make the structure of the action more explicit define
\begin{equation}
\begin{gathered}
R_{ab} = d\omega_{ab}+ \omega_{ac}\omega^c{ }_b  + (e_a f_b - e_b f_a) \,, \qquad T_a=de_a + \omega_{ad}\xi^d  + e_a v \,,
\\
*W^{ab}=W^{abnm}\epsilon_{nmpk}e^p e^k\,,
\qquad *C^a=C^a{ }_{bc}\epsilon^{bcpk}e_p e_k\,.
\end{gathered}
\end{equation}
In these terms the action reads as:
\begin{equation} \label{Cond_form_act}
    S = \int_X \big[R_{ab} \wedge * W^{ab} + W_{ab} \wedge * W^{ab} - 2T_a \wedge *C^a\big] \,.
\end{equation}

\subsection{Symplectic quotient and BV formulation}

In order to give a consistent interpretation to the gauge theory determined by~\eqref{Act} and to construct the respective BV description we need to identify a kernel distribution for the symplectic structure $\bar\omega$, induced on the (open domain of) the space of supermaps $Smaps(\fR^4[1],F)$. In the case at hand the manifold $\bar F$ is coordinatized by the degree zero coordinates introduced in \eqref{fields} together with the remaining coordinates introduced as e.g.:
\begin{equation}
    \hat\sigma^*(\xi^a)=\sum_{l=0}^4 \frac{1}{l!}
    \st{l}{\xi}{}^a_{\mu_1\ldots \mu_l} \theta^{\mu_1}\ldots \theta^{\mu_l} \,,
\end{equation}
with the identification $\st{1}{\xi}{}^a_\mu=e^a_\mu$. The ghost degree of the coefficients is determined by $\gh{\theta^\mu}=1$. We then take $\bar F$ to be the space of supermaps with nondegenerate component $e^a_\mu$ because as we are going to see these coordinates correspond to the frame field which must be invertible.\footnote{This step is completely analogous to the respective considerations~\cite{Grigoriev:2020xec} in the case of Einstein gravity.}

We have the following 
\begin{prop} \label{prop:regul}
The presymplectic structure $\bar \omega$ is regular (i.e. has a constant rank) on $\bar F$.
\end{prop}
The idea of the proof is taken from~\cite{Grigoriev:2020xec} and amounts to identifying vector fields $X_{\alpha}$ on $F$ such that they are in the kernel of $\omega$  and their prolongations to $\bar F$ generate the kernel at any point of the body of $\bar F$. In terms of local coordinates the prolongation of a vector field $X$ on $F$ can be defined in local coordinates through:
\begin{equation}
    \bar{X} \psi^A(\theta) = X^A(\psi(\theta))\,, \qquad X^A(\psi)=X\psi^A\,,
\end{equation}
where in the left hand side $\bar {X}$ acts on the coordinates on $\bar F$, which are packed in the generating function $\psi^A(\theta)$. It is easy to check that $i_X\omega=0$ implies $i_{\bar X} \bar\omega = 0$.

The rank of $\bar\omega$ can't decrease in a neigbourhood of a point. Because the rank of the distribution generated by $\bar X_\alpha$ also can not decrease one concludes that the rank is constant and the kernel distribution is generated by $\bar X_\alpha$. Explicit form of $X_\alpha$ and the proof that their prolongations exhaust the kernel at any point of the body of $\bar F$ is given in the Appendix \ref{sec:kernel}. 

Because $\bar\omega$ is regular the general construction
of Section~\bref{sec:presymp-AKSZ} applies and we end up with the first order  BV formulation whose field-antifield space is $\bar G$ which is a symplectic quotient of $\bar F$. This BV formulation has \eqref{Act} as a classical action. To make sure we are indeed dealing with a specific BV formulation of the conformal gravity we need to make sure that (i) \eqref{Act} reduced to the symplectic quotient is equivalent to the conformal gravity and that (ii) the gauge generators encoded in the BV action take all the gauge symmetries of system into account.

(i) Because \eqref{Act} depends on fields of ghost degree zero only it is enough to analyze the reduction only in the sector of degree zero variables. First of all we disregard all the fields associated to higher order $W$  as \eqref{Act} and $\bar\omega$ do not involve them and hence they can be set to zero. The same applies to $v_\mu$ and $f^a{ }_\mu$. Among $X_\alpha$ the only vector fields that affect the remaining ghost ghost degree 0 variables are
\begin{equation}
\begin{gathered}
X^{(trace)} = \xi^a\frac{\partial}{\partial \xi_{a}} - 2 W^{abnm}\frac{\partial}{\partial W^{abnm}} - 3 C^{abc}\frac{\partial}{\partial C^{abc}}\,,\\
        Y^{(antisym)}_d = \epsilon_{abcd}\xi^c \frac{\partial}{\partial \rho_{ab}} + W_{dnm}{ }^a\epsilon^{nmbc}\frac{\partial}{\partial C^{abc}} \,,\\
    Y^{(trace)}_d = \xi^c \delta^{ab}_{cd} \frac{\partial}{\partial \rho_{ab}} - W^{adbc}\frac{\partial}{\partial C^{abc}}\,. 
\end{gathered}
\end{equation}
The action of their prolongations on $e_{a\mu},\omega_{ab,\mu}$ read as:
\begin{equation}
\begin{gathered}
\bar X^{(trace)} e^a{ }_\mu=e^a{ }_\mu\,,\\
\bar Y^{(antisym)}_d \omega_{ab,\mu}= \epsilon_{abcd}e^c{ }_{\mu}\,,  \qquad
\bar Y^{(trace)}_d \omega_{ab,\mu} = (e_{a,\mu} \eta_{bc} - e_{b,\mu}\eta_{ac})\,.
\end{gathered}
\end{equation}
It follows that by choosing the embedding of $\bar G\subset \bar F$ one can assume that the determinant of $e^a_\mu$ as well as totally antisymmetric and the trace components of $\omega_{ab,\mu}e_c{ }^\mu$ can be set to whatever we want. We use this to set $det(e)=1$ and to set the respective componets of the torsion $T_a=de_a+\omega_{ab} e^b $ to zero. The remaining irreducible component is associated to a hook-type Young tableaux and is set to zero by the equation of motion obtained by varying~\eqref{Act} with respect to $C_{abc}$. Eliminating the remaining component of $\omega$ together with $C_{abc}$ sets $\omega$ to the Levi-Civita connection $\omega_{ab,\mu}(e)$. 
Finally, equations for $W_{abcd}$ can be solved with respect to $W_{abcd}$ itself and its elimination gives:
\begin{equation}
    S[e,\omega(e)] = \int W_{ab} \wedge * W^{ab}\,,
\end{equation}
i.e. the standard action of GGR written in terms of the frame field. The standard argument then implies that all the components of $e^a_\mu$ which do not contribute to the metric can be gauge-fixed by local Lorentz transformations (which are algebraic) and the action can be equivalently rewritten in terms of the metric, giving the proof that we indeed obtained CGR action from the presymplectic AKSZ.

(ii) To check that the BV formulation indeed takes into account all the gauge symmetries of the classical action let us analyze the spectrum of ghost variables.  Ghosts $\lambda,\kappa^a$ are in the kernel and do not survive on $\bar G$. At the same time $\xi^a,\rho^{ab}$ are not affected by the kernel and hence their restriction to the symplectic quotient remain independent coordinates. Their associated gauge transformations are precisely the diffeomorphisms and the local Lorentz transformations. Note that we do not see ghosts associated to Weyl transformations because this invariance is taken into account in this formulation through the kernel of the presymplectic structure and we fixed it when passing to the symplectic quotient. Note also  that there are no ghosts pertaining to the special conformal sector since they are also in the kernel.  We then conclude that we have indeed reconstructed a first order BV formulation of CGR. 

A substantial difference between the presymplectic AKSZ formulation of CGR and the Einstein gravity is that in the case of Einstein gravity one can avoid considering the entire minimal model as a target space and truncate it to just a Poincar\'e algebra with shifted degree, see ~\cite{Grigoriev:2020xec} for more details. The analogous truncation doesn't work in the case of CGR. However, one can still construct a presymplectic AKSZ formulation based on conformal algebra and closely related to the frame-like formulation from~\cite{Kaku:1977pa} but this requires imposing additional constraints by hands. This possibility is discussed in Appendix~\bref{sec:Kaku-str}.

Let us mention that all the coordinates on $(F,q)$ entering the action are irreducible tensors of $O(4,2)$. Moreover, each field can be decomposed into a sum of its self-dual and anti-self-dual components (this would require complexified fields or changing a spacetime signature) which are irreducible under $o(4.2)$. In particular the action can be easily rewritten using double-spinor notations in which the decomposition into the self-dual and anti-self-dual components is manifest. By setting to zero e.g. anti-self-dual component of $W_{abcd}$ and dropping $W^2$-term one arrives to an equivalent representation of the chiral (self-dual) conformal gravity~\cite{Berkovits:2004jj}, see also \cite{Adamo:2016ple,Hahnel:2016ihf,Krasnov:2021nsq}.\footnote{We are grateful to E.~Skvortsov for attracting our attention to the self-dual version of CGR.}

Let us also note that the above presymplectic formulation should admit extension to the case of conformal supergravity, where the underlying algebra is the superconformal algebra~\cite{Kaku:1977pa}. This superalgebra  has eight fermionic generators $Q_\alpha, \Sigma_\alpha$ alongside with the 15 conformal generators $J, P, K, D$ and the axial charge generator $U$.  We denote by $R(\cdot)$ the component of curvature associated to the respective generator, i.e. $R(P_a) = T_a = de_a + \omega_{ab}e^b + e_a \lambda $. Then we introduce auxiliary fields associated to the components of the curvature that are nonzero in the respective minimal model. In the CGR case these were $W_{ab}(J_{cd})=W_{abcd}$ and $W_{ab}(K_c)=C_{abc}$. In the case of supergravity the nonvanishing curvatures can be infered from~\cite{Fradkin:1985am} so that one in addition has the following auxiliaries: $W(Q)$, $W(\Sigma)$, $W(U)$. By inspecting our presymplectic action~\eqref{Cond_form_act} it is clear that it has a natural extension of the same structure. More precisely, it reads schematically as 
\begin{multline*}
R(J)\wedge *W(J) +R(U)\wedge *W(U) + \\ +R(P)\wedge *W(K)+ R(Q)\wedge *W(\Sigma)+R(\Sigma)\wedge *W(Q) +\\ +W(J)\wedge *W(J) +W(U)\wedge *W(U)+W(Q)\wedge *W(
\Sigma)\,,
\end{multline*}
where the pairing is determined by the invariant inner product on the superconformal algebra. Of course, determining such an action is equivalent to determining an underlying presymplectic form. We expect that the action of this structure should do the job but refrain from making a precise statement because it might involve extra constraints and gauge-fixings.

\section{General presymplectic BV-AKSZ formalism}
\label{sec:gen-presymp}

Our discussion in Section~\bref{sec:AKSZ-1} was limited to the case of diffeomorphism-invariant systems for which AKSZ representation (at the level of EOMs) is available. Now we explain how the presymplectic BV-AKSZ formalism extends to the general situation  of not-necessarily diffeomorphism-invariant systems.

\subsection{Gauge PDEs}

If we are only interested in the equations of motion and gauge symmetries (and hence disregard the Lagrangian) the data of a local gauge theory can be encoded into the following geometric object, known as gauge PDE \cite{Grigoriev:2019ojp} (see also \cite{Barnich:2010sw,Grigoriev:2010ic,Grigoriev:2012xg} for the earlier and less-geometrical version):
\begin{definition}
Gauge PDE $(E,T[1]X,Q)$ is a $\fZ$-graded fiber bundle $\pi: E\to T[1]X$  whose total space is equipped with a homological vector field $Q$ of degree $1$ satisfying: $Q \circ \pi^*=\pi^* \circ d_X$
and $d_X$ is the de Rham differential on $X$ seen as a homological vector field on $T[1]X$. Moreover, $(E,Q)$ should be equivalent to a non-negatively graded gauge PDE.
\end{definition}
One also often imposes some extra conditions of a technical nature to exclude pathological examples.
We assume that $X$ is a real manifold (space-time manifold) while $E$ (and $T[1]X$) is a $\fZ$-graded one with the degree denoted by $\gh{\cdot}$. Note that gauge PDEs are special cases of $Q$-bundles~\cite{Kotov:2007nr}. However, gauge PDEs are necessarily $\fZ$-graded and are typically infinite-dimensional. Further details and references can be found in~\cite{Grigoriev:2019ojp}.

Just like in the case of AKSZ sigma models discussed in Section~\bref{sec:AKSZ-1}, solutions of a gauge PDE are $Q$-sections, i.e. sections satisfying $\sigma^*\circ Q=d_X \circ \sigma^*$, cf. \eqref{aksz-eom}. Infinitesimal gauge transformation of sections is also defined in exactly the same way through the  relation~\eqref{aksz-gs}.

The notion of gauge PDE is rather flexible. For instance, the standard BV formulation at the level of equations of motion is reproduced by taking as $E$ a jet-bundle associated to the bundle of BV fields (fields, ghosts, antifields) seen as a bundle over $T[1]X$ (i.e. functions on $E$ are local horizontal forms on the jet-bundle) and $Q=\dh+s$, where $\dh$ is a canonical horizontal differential on the jet-bundle and $s$ is the BV-BRST differential. It was shown in~\cite{Barnich:2004cr} that, at least locally,  this  gauge PDE gives an equivalent (in the sense of elimination of generalized auxiliary fields) description of the same gauge system.

Another standard example is PDE. Let $E_X \to X$ be an equation manifold seen as a bundle over the space-time $X$. The Cartan distribution gives rise to a homological vector field $\dh$ on $E$ which is $E_X$  pulled back to $T[1]X$. In a suitable local coordinate system $x^a,\theta^a,\psi^A$, where $x^a, \theta^a$ are adapted local coordinates on $T[1]X$ , induced by local coordinates $x^a$ on $X$, and $\psi^A$ originate from local coordinates on a typical fiber, one takes:
\begin{equation}
    \dh=\theta^a D_a\,, \qquad D_a=\dl{x^a}+\Gamma_a^A(x,\psi)\dl{\psi^A}\,,
\end{equation}
where $D_a$ are components of the total derivative on the equation manifold. Note that this is a generic form of the $Q$-structure on $E$ if the typical fiber is a real manifold (so that $\gh{\psi^A}=0$). It is easy to see that in these coordinates solutions are covariantly-constant sections in agreement with the standard picture. Mention also that a closely related representation of PDEs is known under the name of unfolded formulation~\cite{Vasiliev:1988xc,Vasiliev:2005zu}.

From the gauge PDE perspective AKSZ sigma models discussed in Section~\bref{sec:AKSZ-1} correspond to trivial (as a $Q$-bundle) $Q$-bundles $(E,Q)=(F,q)\times (T[1]X,d_X)$, where $(F,q)$ is a fiber.

\subsection{Vertical forms}

The Lagrangian (or partially Lagrangian) systems can be described as gauge PDEs with extra structures. To define these we first need to recall a particular realization of vertical forms.

Let $\pi:E \to \manX$ be a generic fiber bundle. The exterior algebra of local forms on $E$ contains the ideal $\cI$ generated by forms of the form $\pi^*\alpha$, where $\alpha \in \bigwedge^k (\manX)$, $k>0$. The algebra of vertical forms is the quotient of $\bigwedge^{\bullet} (E)$ by the ideal.  Note that vertical zero-forms are just zero-forms as the ideal does not contain zero-forms. It is convenient to work with vertical forms in terms of the representatives as we do in what follows.

The De Rham differential $d$ is well defined on the equivalence classes. Understood as an operation on vertical forms it is usually refereed to as a vertical differential. Suppose that $Y$ is a vector field on $E$, which is related by the projection to a vector field $y$ on $\manX$, i.e. $Y\circ \pi^* =\pi^* \circ y$, which in turn implies $L_Y\circ \pi^* =\pi^* \circ L_y$. It is easy to check that $L_Y$ is well defined on the equivalence classes and hence defines a Lie derivative on vertical forms. Despite $i_Y$ is not well-defined on equivalence classes (unless $Y$ is vertical) one can still employ the Cartan formula $L_Y=i_Y d+(-1)^{\p{Y}} d i_Y$ for representatives.

Given a local trivialization of $E$ over $U\subset \manX$, i.e. $E|_{U}=U\times F$, each equivalence class has a unique representative that vanishes on any horizontal vector. In the  adapted coordinates $\psi^A,X^\alpha$, with $X^\alpha$ being coordinates on the base, this representative does not involve basis differential $dX^\alpha$.  In terms of such representatives and the adapted coordinates the vertical differential can be explicitly written as $d\psi^A\dl{\psi^A}$. It is then clear that the vertical differential is locally  acyclic in positive form-degree.

\subsection{Presymplectic gauge PDEs}

A presymplectic BV-AKSZ system is a gauge PDE $(E,T[1]X,Q)$  equipped with a compatible vertical presymplectic structure $[\omega]$, i.e. a 2-form $\omega$ satisfying $\gh{\omega}=n-1$, $d\omega \in \cI$, $L_Q\omega=\cI$, where $n=\dim{X}$. In addition we assume the following technical conditions: $i_Qd\omega\in \cI$ and $i_Q L_Q\omega\in \cI$. Here and in what follows we work in terms of representatives.

Given a presymplectic gauge PDE one can consider a natural  Lagrangian system associated to it. More precisely, we first introduce covariant Hamiltonian $\cL$ as a solution to\footnote{In the case of constrained mechanical systems $\hL=\Omega-\theta H$,
where $H$ is a Hamiltonian and $\Omega$ is the BRST charge. 
In this context the formulation in question and $\cL$, in particular, was proposed  in~\cite{Grigoriev:1999qz} (see also~\cite{Batalin:1998pz,Batalin:1997ks} where $\hL$ appeared earlier in a slightly different context).}
\begin{equation}
\label{H:def}
i_Q\omega + d\hL \in \cI\,.
\end{equation}
To see that $\hL$ exists observe that $d i_Q\omega \in \cI$ thanks to $L_Q\omega\in \cI$ and $i_Q d\omega \in \cI$. Then acyclicity of $d$ on vertical forms implies the existence of $\hL$ (for $n>1$ it exists globally). Note that $\hL$ is a zero form and hence  is defined uniquely modulo functions of the form $\pi^*(h)$, $h\in\CC^\infty(T[1]X)$. As we are going to see this ambiguity doesn't affect the equations of motion. 

If we pick a symplectic potential $\chi$ satisfying $\omega-d\chi\in \cI$ this data defines a natural action functional on sections $T[1]X \to E$:
\begin{equation}
S[\sigma]=\int_{T[1]X} [(\sigma^*\,\chi)(d_X)+\sigma^*(\hL)]\,.
\end{equation}
This action is a generalization of the presymplectic version~\cite{Alkalaev:2013hta} of the conventional AKSZ action discussed in Section~\bref{sec:AKSZ-1}. At the same time, in the case where $(E,T[1]X,Q)$ is a PDE, it coincides with the so-called intrinsic action from~\cite{Grigoriev:2016wmk} (see also~\cite{Grigoriev:2021wgw}). In the later case $\omega$ can be identified with a compatible presymplectic current (see~\cite{Kijowski:1979dj,Crnkovic:1986ex,Zuckerman:1989cx,Anderson1991,Khavkine2012} for further details) defined on the equation manifold. 

Let us check that $S[\sigma]$ does not depend on the choice of a representative $\chi$ of the equivalence class $[\chi]$. If instead of $\chi$ we took another representative $\chi^\prime=\chi+\epsilon$, where $\epsilon \in \cI$ so that it can be written in local coordinates as $\epsilon =dx^a \alpha_a +d \theta^b \beta_b $ we find:
\begin{equation}
(\sigma^*(\epsilon))(d_X)=\theta^a \sigma^*(\alpha_a)\,.
\end{equation}
At the same time, the variation $\hL_\epsilon$ of $\hL$ is determined by $i_Q d\epsilon+d\hL_\epsilon\in \cI$. Using $L_Q\epsilon \in \cI$ one finds $di_Q\epsilon + d\hL_\epsilon \in \cI$ and hence $\hL_\epsilon+i_Q\epsilon=\pi^*(h)$.  A variation of the form $\pi^*(h)$, $h\in \CC^\infty(T[1]X), \gh{h}=n$
is locally $d_X$-exact and hence it preserves the equivalence class of the Lagrangian. Furthermore,  
\begin{equation}
\sigma^*(i_Q \epsilon)=\sigma^*(\theta^a \alpha_a)=\theta^a\sigma^*(\alpha_a)\,,
\end{equation}
where we again use adapted coordinate system $x^a,\theta^a,\psi^A$ on $E$. Note that  in this coordinates $Qx^a=\theta^a$, $Q\theta^a=0$. It follows that the action is unchanged under $\chi \to \chi^\prime$ if we disregard the variations of the form $\pi^*(h)$, which do not affect equations of motion.

In this way we have arrived at a natural Lagrangian system associated to a presymplectic BV-AKSZ system. However, in general this system contains an infinite amount of fields and the interpretation of the action is to be clarified. In the case where  $(E,T[1]X,Q)$ is a usual PDE (seen as a gauge PDE) the consistent interpretation can be arrived at by disregarding those fields which are in the kernel of the symplectic structure~\cite{Grigoriev:2016bzl}. In this way one can often (though not always) describe Lagrangian systems in terms of the intrinsic geometry of their equation manifolds. Just like in the case of diff-invariant systems discussed in Section~\bref{sec:AKSZ-1} a more refined interpretation can be given by passing to the symplectic quotient of the induced symplectic structure  on the space of super-sections of $E$. 

\subsection{Symplectic quotient and BV formulation}

It is convenient to identify super-sections as the space of section of a bundle $\bar E$ over $X$, whose fiber over $p\in X$ is the space of supermaps from $(T_pX)[1]$ to $F$.

To be more explicit we now work locally using a trivialization $E|_{T[1]U}=T[1]U\times F$ over $U\subset X$.  The fiber $\bar F$ of $\bar E|_U$ is  a space of supermaps from $\fR^{n}[1]$ to $F$. Coordinates on $\bar F$ are introduced as in \eqref{scoord}, where we keep denoting coordinates on $F$ by $\psi^A$ and use $\Psi^I$ to denote coordinates on $\bar F$.

It turns out that the presymplectic structure $\omega$ on $E$ gives rise to an $n+2$-form on $\bar E$ of the following structure:
\begin{equation}
\bar\omega^{\bar E}=(dx)^n \bar \omega_{IJ}(\Psi,x) d\Psi^I d\Psi^J\,,
\end{equation}
where $\bar\omega_{IJ}=\bar \omega(\dl{\Psi^I},\dl{\Psi^J})$, with $\bar \omega$ on each fiber given by
\begin{equation}
    \bar\omega=\int d^n\theta \omega_{AB}(\psi^A(\theta),x,\theta) d\psi^A(\theta)\, d\psi^B(\theta)\,.
\end{equation}
In other words on a given fiber the form is still the same~\eqref{bar-omega} tensored with the coordinate volume form $(dx)^n$.
It is easy to see that $d\bar\omega=0$. An alternative language would be to identify $\bar\omega$ as a vertical 2-form with values in densities on $X$.\footnote{A more geometrical way to define $\bar\omega$  is to introduce super-jet bundle of $E\to T[1]X$ and pullback $\omega$ to the super-jet bundle. In a suitable coordinate system this can be seen as a term in $\bar\omega$ proportional to $(\theta)^n$.}

We assume that the $n+2$ form $\bar\omega$ is regular (in the sense that the rank of $\bar\omega_{IJ}$ is constant) and consider the vertical kernel of $\bar\omega$. This consists of vertical vector fields $W$ such that $i_W \bar\omega=0$. Thanks to $d\bar\omega=0$ the vertical kernel is an involutive distribution and one can (at least locally) pass to the quotient space. Because the kernel distribution is vertical the quotient is again a bundle over $X$ (its fibers are simply the symplectic quotients of the symplectic form $\bar\omega_{IJ}d\Psi^i d\Psi^J$ on $\bar F$). Form $\bar\omega$ induces a vertically-nondegenerate $n+2$-form on the quotient. This gives us a fiber-bundle equipped with the $n+2$-form satisfying all the conditions of the BV $n+2$-form. It has a natural lift to an $(n,2)$-form on $J^\infty(\bar E)$.

By repeating the arguments given in Section~\bref{sec:presymp-AKSZ} in this more general setting one can show that BV-like action \eqref{BV-like} still induces the BV action on the symplectic quotient and it satisfies the BV master equation modulo boundary terms. 

To summarize: under some regularity assumptions a presymplectic gauge PDE in a natural way encodes a Lagrangian BV system. This system is not necessarily equivalent (at the level of equations of motion) to the starting point gauge PDE. Neveretheless, for a given Lagrangian BV system one can always find a presymplecic gauge PDE such that the BV system it encodes is equivalent to the initial BV system. More precisely, this can be done by using the Lagrangian parent formulation~\cite{Grigoriev:2010ic} as explained in~\cite{Grigoriev:2016wmk,Grigoriev:2021wgw}. Note, however, that there exist equivalent gauge PDEs such that one of them admits a compatible presymplectic structure leading to an equivalent BV system while another one does not. The simplest example is provided by the Fierz-Pauli Lagrangian describing massive spin-2 field. It turns out~\cite{Grigoriev:2021wgw} that the respective equation manifold does not admit a compatible presymplectic structure that reproduces an equivalent Lagrangian.

\subsection{Example: Maxwell}
We now illustrate the construction using the example of Maxwell theory. More precisely we take the minimal model of the respective gauge PDE (the respective total space $Q$-manifold is known in the literature for quite some time, see e.g.~\cite{Brandt:1996mh} and refs. therein. The solutions of this gauge PDE satisfy the so-called unfolded form of Maxwell equations, see e.g.~\cite{Lopatin:1987hz}). The coordinates on the total space of $E\to T[1]X$ are 
\begin{equation}
\begin{gathered}
x^a,\theta^a,\quad C\,,\quad \gh{C}=1\\
    F_{a|b}, \quad F_{a|bc},\quad \ldots \quad  F_{a|b_1\ldots b_l}\,,\quad \ldots\,,\quad \gh{F_{a|b\ldots}}=0\,,
\end{gathered}
\end{equation}
where  tensors $F_{a|b_1\ldots b_l}$ are symmetric in the second group of indices,  satisfy the Young symmetry condition $F_{(a|b_1\ldots b_l)}=0$, and are totally traceless.  The action of $Q$ is given by 
\begin{equation}
\begin{gathered}
Qx^a=\theta^a\,,\quad QC=\half \theta^a\theta^b F_{a|b}\,,
\\
QF_{a|b}=\theta^c(F_{a|bc}+\ldots)\,, \quad QF_{a|bd}=\theta^c(F_{a|bdc}+\ldots)\,, \quad \ldots  
\end{gathered}
\end{equation}
where $\ldots$ denote terms maintaining the Young symmetry properties. It is clear that we have a $Q$-bundle structure as $Q\circ \pi^*=\pi^*\circ d_X$, with $d_X=\theta^a\dl{x^a}$ and $\pi^*(x^a)=x^a$ and $\pi^*(\theta^a)=\theta^a$ in these coordinates. 

The compatible presymplectic structure and the respective potential can be taken as~\cite{Alkalaev:2013hta}:
\begin{equation}
    \omega=dC dF^{a|b} (\theta)^{(n-2)}_{ab}\,, \qquad \chi = dC F^{a|b} (\theta)^{(n-2)}_{ab}\,,
\end{equation}
where $(\theta)^{(n-k)}_{a_1\ldots a_k}=\frac{1}{(n-k)!}\epsilon_{a_1\ldots a_k b_1\ldots b_{n-k}}\theta^{b_1}\ldots \theta^{b_{n-k}}$. Note that in checking $L_Q\omega \in \cI$ one has to use the "equations of motion", i.e. that $F_{a|bc}$ are totally traceless. Recall that ideal $\cI$ is generated by $dx^a,d\theta^a$.

The covariant Hamiltonian $\hL$ defined through $i_Q \omega+d\hL\in \cI$ is given by
\begin{equation}
    \hL=-\half F_{a|b}F^{a|b}(\theta)^{(n)}
\end{equation}
Introducing component fields parameterizing  section $\sigma:T[1]X\to E$ according to
\begin{equation}
\sigma^*(C)=A_b(x)\theta^b\,, \qquad \sigma^*(F_{a|b})=F_{ab}(x)\,, \qquad \ldots
\end{equation}
the action takes the form
\begin{equation}
\label{1st-maxwell}
S[\sigma]=\int d^n x \big[F^{ab}(\dd_a A_b -\dd_b A_a) - \half F^{ab}F_{ab}\big]\,,
\end{equation}
which is a standard covariant 1-st order action of the Maxwell field. 

The induced symplectic structure $\bar\omega$ on the fiber $\bar F$ is given by 
\begin{equation}
\bar\omega=d\st{0}{C}\, d \st{2}{F}{}^{a|b}_{ab}
+2 dA_a\, d\st{1}{F}{}^{a|b}_{b}
+d \st{2}{C}{}_{ab} \, d\st{0}{F}{}^{a|b} \,,
\end{equation}
where the relevant components of supersection are introduced via:
\begin{equation}
\begin{aligned}
     \hat\sigma^*(C)&={\st{0}{C}(x)}+{A_a(x)} \theta^a+{\half}{{\st{2}{C}_{ab}(x)}}\theta^a\theta^b \ldots\\
     \hat\sigma^*(F^{a|b})&={{F}{}^{a|b}(x)}
     +{\st{1}{F}{}^{a|b}_c (x)}\theta^c
     +\half {\st{2}{F}{}^{a|b}_{cd}(x)}\theta^c\theta^d+\ldots\,.
\end{aligned}
\end{equation}
It follows that all the fields except for 
\begin{equation}
    \st{0}{C},\quad  C^*=\st{2}{F}{}^{a|b}_{ab}, \quad A_a, \quad  A_*^a=\st{1}{F}{}^{a|b}_b, \quad  {F}{}^{a|b},\quad F^*_{a|b}=\st{2}{C}{}_{ab}
\end{equation}
are in the kernel of $\bar\omega$ and can be set to zero to obtain the 
symplectic quotient. The ghost degree of the remaining fields is given by 
\begin{equation}
\begin{gathered}
\gh{A_a}=\gh{F^{a|b}}=0\,, \qquad \gh{C}=1\,,\\ \gh{A^A_*}=\gh{F^*_{a|b}}=-1\,,\qquad \gh{C^*}=-1\,.
\end{gathered}
\end{equation}
By inspecting the degree of the remaining fields and the above symplectic structure one concludes that we have arrived at the proper field-antifield space of the BV formulation of the Maxwell field. It is straightforward to check that the BV action~\eqref{BV-like} specialised to this case explicitly gives a proper solution of the master equation extending the classical action~\eqref{1st-maxwell}. Extension to the YM theory is also straightforward.

As a final remark mention that in 4 dimensions it is not difficult to find a presymplectic BV-AKSZ formulation of the self-dual version of the system. More precisely, all $F$-coordinates can be decomposed into the self dual $F^+$ and anti-self-dual $F^-$ components.  It is easy to check that replacing $F$ with $F^+$ in the relation $QC=\half \theta^a\theta^b F_{a|b}$ and replacing $F$ with $F^-$ in the expression for the presymplectic  structure results in the presymplectic AKSZ formulation of the self-dual Maxwell theory. Nonabelian extension is straightforward and results in the SDYM theory~\cite{Chalmers:1996rq}.

\section{Conclusion}

In this work we have proposed the presymplectic BV-AKSZ formulation of the conformal gravity in 4 dimensions. Among possible formulation of this type the present one is minimal in the sense that it is based on the minimal model of the respective total BRST complex and hence it involves a minimal number of fields among possible AKSZ-type formulations.  In particular, fields of the system are irreducible Lorentz tensors.  In a certain sense the proposed formulation is a "canonical" first-order frame like-formulation. At the same time this does not mean that other formulations can not be more useful. For instance an interesting issue is to extend the present formulation to the one where $o(4,2)$-symmetry is realised in a manifest way.

The proposed formulation could be also useful in describing less symmetric models such as CGR with the extra $R^2$ counterterm added. Although such system is not Weyl invariant and a minimal presymplectic formulation should involve only Poincar\'e
algebra one can try to keep the entire  conformal algebra at the price of introducing the appropriate compensator field. This is very much in spirit of tractor description of Riemannian geometry, see e.g.~\cite{Curry:2014yoa}.

Another interesting open problem is the extension of the approach to the case of conformal higher spin fields~\cite{Fradkin:1985am} (see also~\cite{Fradkin:1989md,Beccaria:2014jxa, Bekaert:2013zya,Kuzenko:2019ill} for a more recent discussion from various perspectives) for which the interacting local theory is known~\cite{Segal:2002gd,Tseytlin:2002gz} (see also \cite{Bekaert:2010ky,Bonezzi:2017mwr}) and present a substantial interest from various perspectives. It would also be desirable to make the relation with the tractor description of conformal geometry~\cite{Cap:2002aj,Curry:2014yoa} more precise.  

In the present work we also extend the approach to not necessarily diffeo\-mor\-phism\--invariant theories. In this case the AKSZ framework is not enough and one needs to resort to so called gauge PDEs of which the usual AKSZ setup is a particular case. Another extreme case is the usual geometrical description of PDE in terms of the equation manifold (for a modern review see e.g.~\cite{Krasil'shchik:2010ij}). In this case the presymplectic BV-AKSZ formulation reduces to the so-called intrinsic Lagrangian approach~\cite{Grigoriev:2016wmk,Grigoriev:2021wgw}. As an illustration of the general situation we consider a minimal model of the Maxwell system endowed with a compatible presymplectic structure and explicitly demonstrate how the full-scale BV description emerges from it.

\section*{Acknowledgments}
\label{sec:Aknowledgements}
ID acknowledges useful discussions with V.~Gritzaenko and A.~Yan. MG wishes to thank N.~Boulanger, A.~Sharapov, E.~Skvortsov and especially  A.~Kotov for fruitful discussions. The work was supported by the Russian Science Foundation grant 18-72-10123.
Part of this work was done when authors participated in the thematic program "Higher Structures and Field Theory" at the Erwin Schrödinger International Institute for Mathematics and Physics, Vienna, Austria. This work was partially completed while M.G. participated in  the workshop "Higher Spin Gravity and its Applications" supported by the Asia Pacific Center for Theoretical Physics, Pohang, Korea.

\appendix
\section{Tensor conventions}\label{sec:techincal}

In this section we describe our conventions for the tensors emerging in the context of CGR and give missing details on its frame-like description.

The Bach tensor is defined as 
\begin{equation}
    B_{\mu \nu} = P^{\rho \lambda}W_{\mu \rho \nu \lambda} + \nabla^\rho \nabla_\rho P_{\mu \nu} - \nabla^\rho \nabla_\mu P_{\nu \rho}\,,
\end{equation}
where $\nabla^\mu$ is the metric-compatible covariant derivative, $P_{\mu \nu}$ is the Schouten tensor given by
\begin{equation}
    P_{\mu \nu} = \frac{1}{2}(\stackrel{(0)}{R}{}^\rho{ }_{\mu\rho \mu} - \frac{\stackrel{(0)}{R}{}^{\rho \sigma}{ }_{\rho \sigma}}{6} g_{\mu \nu})\,,
\end{equation}
$\stackrel{(0)}{R}{}^\rho{ }_{\mu\nu\sigma}$ is the Riemann curvature, and Weyl tensor is defined as:
\begin{equation}
    W_{\mu \nu}{ }^{\rho \sigma} = \stackrel{(0)}{R}{}_{\mu \nu }{ }^{\rho \sigma} - 4 P_{[\mu}^{[\rho}\delta_{\nu]}^{\sigma]}\,.
\end{equation}

Now we give missing details on the derivation of the metric-like action~\eqref{metric-action} from the frame-like action~\eqref{action}. Starting with the equation of motion $\epsilon^{abcd} R_{ab} e_c = 0$ for $f^a$, introducing tangent components $R_{abcd}$ via $R_{ab} = R_{abcd}e^c e^d$, and wedging with $e^f$  one finds 
\begin{equation}
    (R^{ab}{ }_{km})\delta^{k}_{[a} \delta^m_b \delta_{d]}^f = 0\,,
\end{equation}
which implies
\begin{equation}
    2R^{ab}{ }_{ab}\delta^f_d + 4R^{fb}{ }_{bd} = 0\,.
\end{equation}
Contracting $f,d$ one gets $R^{ab}{ }_{ab} = 0$ and 
\begin{equation}\label{EOMfAp}
    R^c{ }_{bcd} = 0\,.
\end{equation}
Explicitly one has
\begin{equation}
    R^c{ }_{bcd} = \stackrel{(0)}{R^c{ }_{bcd}} + (4f_{d,b} - e_{d,a} f_{a,b} - e_{a,b} f_{d,a} + e_{d,b}f) = 0\,,
\end{equation}
where $f = f^a_a$. It follows 
\begin{equation}
    f = -\frac{1}{6}\stackrel{(0)}{R}
\end{equation}
and
\begin{equation} \label{SchoutenAp}
    f_{b,\mu} = -\frac{1}{2}(e^d{ }_{\mu}\stackrel{(0)}{R^c{ }_{bcd}} - \frac{1}{6}e_{b,\mu}\stackrel{(0)}{R})
\end{equation}
so that $f_\mu^b$ is indeed a Schouten tensor on shell. 

Now we turn to the properties of the Weyl and Cotton tensors and their Hodge duals, which we need in the proof of the invariance of the presymplectic structure. The Hodge duals are defined as:
\begin{equation}
    C^*_{abc} = C_a{ }^{pk}\epsilon_{pkbc} \,, \qquad 
    W^*_{abcd} = W_{ab}{ }^{pk}\epsilon_{pkcd}\,.
\end{equation}
The inverse relations read:
\begin{equation}
\begin{aligned}
       C^*_{abc}\epsilon^{bcnm} &= C_a{ }^{pk}\epsilon_{pkbc}\epsilon^{bcnm} = 4C_a{ }^{nm}\,,\\
    W^*_{abcd}\epsilon^{cdnm} &= W_{ab}{ }^{pk}\epsilon_{pkcd}\epsilon^{cdnm} = 4 W_{ab}{ }^{nm}\,.
\end{aligned}
\end{equation}    
Every property of Weyl and Cotton tensors has its counterpart in terms of the duals. 
\begin{enumerate}
\item Trace-free and Young symmetry
\begin{equation} \label{Ctrpr}
\begin{gathered}
    \eta^{ac} C_{abc} = 0 = {C^*_{apk}\epsilon^{pkba}}\,, \\
    \epsilon^{abcd} C_{abc} = 0 =  C^*_{a}{ }^{pk}\epsilon_{pkbc}\epsilon^{abcd} = 2C^*_{a}{ }^{ad}\,,
\end{gathered}
\end{equation}
\begin{equation} \label{Wtrpr}
\begin{gathered}
    \eta^{ac} W_{abcd} = 0 = {W^*_{abpk}\epsilon^{pkad}}\,, \\
    \epsilon^{abcn} W_{abcd} = 0 =  W^*_{ab}{ }^{pk}\epsilon_{pkcd}\epsilon^{abcn} = 2W^*_{ab}{ }^{ab}\delta^n_d + 8W^*_{ad}{ }^{an}\,.
\end{gathered}
\end{equation}

\item Bach equation 
\begin{equation} \label{CBach}
    C_{abc|}{ }^b = 0 = {C^*_{a}{ }^{pk|b}\epsilon_{pkbc}}\,.
\end{equation}

\item Definition of $C$ in terms of higher Weyl-tensors:
\begin{equation} \label{WCrel}
     W_{abcd}{ }^{|a} = W_{bcda}{ }^{|a} -  W_{bdca}{ }^{|a} = C_{bcd}\,.
\end{equation}

\item The Weyl tensor enjoys the following relation: 
\begin{equation} \label{Wchpr}
    W_{ab}{ }^{ef}\epsilon_{cdef} = W^{ef}{ }_{cd}\epsilon_{abef}\,.
\end{equation}

\end{enumerate}

\section{Invariance of the presymplectic structure} \label{sec:presymp-details}
Here we explicitly demonstrate that the presymplectic structure \eqref{presymp-main} is $\tgamma$-invariant.

Equation \eqref{iqomw} is obtained in the following way:
\begin{multline}
    i_{q}\omega_W = \rho_{al}\rho^l{ }_b d(W^{abnm}\epsilon_{nmpk}\xi^p\xi^k) + (\xi_a\kappa_b - \xi_b\kappa_a)d(W^{abnm}\epsilon_{nmpk}\xi^p\xi^k) +\\+ \frac{1}{2}W_{abij}\xi^i\xi^jd(W^{abnm}\epsilon_{nmpk}\xi^p\xi^k)+ \\+ d\rho_{ab}(\xi^j  W^{abnm}{ }_{|j} + \rho^a{ }_j W^{jbnm} +\rho^b{ }_j W^{ajnm} + \rho^n{ }_j W^{abjm} + \\ + \rho^m{ }_j W^{abnj} + 2\lambda W^{abnm})\epsilon_{nmpk}\xi^p\xi^k + \\ + 2d\rho_{ab}W^{abnm}\epsilon_{nmpk}(\rho^p{ }_j \xi^j\xi^k + \xi^p\lambda \xi^k) = \\ = d(\rho_{al}\rho^l{ }_b W^{abnm}\epsilon_{nmpk}\xi^p\xi^k) + (\xi_a\kappa_b - \xi_b\kappa_a)d(W^{abnm}\epsilon_{nmpk}\xi^p\xi^k) + \\ + \frac{1}{2}W_{abij}\xi^i\xi^j d(W^{abnm}\epsilon_{nmpk}\xi^p\xi^k) + d(\rho_{ab})\xi^j W^{abnm}{ }_{|j}\epsilon_{nmpk}\xi^p\xi^k\,,
\end{multline}
where we made use of:
\begin{multline}
\rho_{al}\rho^l{ }_b d(W^{abnm}\epsilon_{nmpk}\xi^p\xi^k) + d\rho_{ab}(\rho^a{ }_j W^{jbnm} +\rho^b{ }_j W^{ajnm})\epsilon_{nmpk}\xi^p\xi^k =\\= d(\rho_{al}\rho^l{ }_b W^{abnm}\epsilon_{nmpk}\xi^p\xi^k)\,.
\end{multline}

For the second piece $\omega_C$ of the presymplectic structure we have
\begin{multline}
    i_{\tgamma} \omega_{C} = \rho_{an}\xi^n d(C^a{ }_{bc}\epsilon^{bcpk}\xi_p \xi_k) + \xi_a \lambda d(C^a{ }_{bc}\epsilon^{bcpk}\xi_p \xi_k) +\\+ d(\xi_a)(\xi^d D_d C_{abc} + \rho_a{  }^n C_{nbc} + \rho_b{  }^n C_{anc} + \rho_c{  }^n C_{abn} + 3\lambda C_{abc}  + \kappa_n W^n{ }_{abc} )\epsilon^{bcpk}\xi_p \xi_k + \\ + 2 d(\xi_a)C^a{ }_{bc}\epsilon^{bcpk}(\rho_{p}{ }^l \xi_l \xi_k + \xi_p \lambda \xi_k)\,,
\end{multline}
where the following cancellations can be observed:
\begin{equation}
3d(\xi_a)\lambda C_{abc}\epsilon^{bcpk}\xi_p \xi_k + 2 d(\xi_a)C^a{ }_{bc}\epsilon^{bcpk}\xi_p \lambda \xi_k = d(\xi_a)\lambda C_{abc}\epsilon^{bcpk}\xi_p \xi_k 
\end{equation}
and 
\begin{multline}
d(\xi_a)(\rho_b{  }^n C_{anc} + \rho_c{  }^n C_{abn})\epsilon^{bcpk}\xi_p \xi_k + 2 d(\xi_a)C^a{ }_{bc}\epsilon^{bcpk}\rho_{p}{ }^l \xi_l \xi_k = 0\,.
\end{multline}
Taking them into account we indeed arrive at
\eqref{iqomc}. To see that terms from equations \eqref{Wdisc},\eqref{Cdisc} give \eqref{finres} we use properties \eqref{Wtrpr}, \eqref{Ctrpr}, \eqref{CBach}, \eqref{WCrel}, \eqref{Wchpr}. The nontrivial calculation is to show that:
\begin{equation}
    {d(\rho^{bc})\xi^a \xi^p \xi^k  W^*_{bcpk|a} = -2d(\rho^{bc})\xi_c \xi^p \xi^k C^*_{bpk}}.
\end{equation}
To prove it we start with \eqref{WCrel}:
\begin{equation}
\begin{gathered}
    W_{abcd}{ }^{|a} = W_{bcda}{ }^{|a} -  W_{bdca}{ }^{|a} = C_{bcd}\,, \\ 
    W_{cdba}{ }^{|a} - W_{cbda}{ }^{|a} = C_{cdb}\,, \\
    W_{dbca}{ }^{|a} - W_{dcba}{ }^{|a} = C_{dbc}\,,
\end{gathered}
\end{equation}
\begin{equation}
     W_{bcda}{ }^{|a} = \frac{1}{2}(C_{bcd} + C_{cdb} - C_{dbc} ) = (C_{bcd} + C_{cdb})\,,
\end{equation}
\begin{equation} \label{WCreld}
\begin{gathered}
    W^*_{bc}{ }^{pk|a} \epsilon_{pkda} =( C^*_{b}{ }^{pk} \epsilon_{pkcd} +  C^*_{c}{ }^{pk} \epsilon_{pkdb}) \,.
\end{gathered}
\end{equation}
Multiply this equation by $\epsilon^{dnmr}$ to get
\begin{equation}
    \delta^{nmr}_{pka}W^*_{bc}{ }^{pk|a} = (-C^*_{b}{ }^{pk} \delta^{nmr}_{pkc} + C^*_{c}{ }^{pk}\delta^{nmr}_{pkb} )\,,
\end{equation}
\begin{equation}
    \xi^a \xi^p \xi^k  W^*_{bcpk|a} = (-\xi_c \xi^p \xi^k C^*_{bpk} + \xi_b \xi^p \xi^k C^*_{cpk})\,.
\end{equation}
Contracting this equation with $d\rho^{bc}$ yields the desired relation.

We conclude that 
\begin{multline} \label{invcheckAp}
    i_{\tgamma}(\omega_W - 2\omega_{C}) = d(\rho_{al}\rho^l{ }_b W^{abnm}\epsilon_{nmpk}\xi^p\xi^k) + d(2\xi_a\kappa_b W^{abnm}\epsilon_{nmpk}\xi^p\xi^k) + \\  + d(-2\rho_{an}\xi^n C^a{ }_{bc}\epsilon^{bcpk}\xi_p \xi_k) + d(-2\xi_a \lambda C^a{ }_{bc}\epsilon^{bcpk}\xi_p \xi_k) + \\+ \frac{1}{2}W_{abij}\xi^i\xi^j d(W^{abnm}\epsilon_{nmpk}\xi^p\xi^k)\,.
\end{multline}

\section{The kernel of the presymplectic structure}  \label{sec:kernel}

Here we study the kernel of the presymplectic structure~\eqref{presymp-main} and show the properties needed in the proof of Proposition~\bref{prop:regul}. We first identify
explicitly vector fields on $F$ which are in the kernel of $\omega$ and whose prolongations generate the kernel distribution for $\bar\omega$ on $\bar F$. It is convenient to group them as follows:
\begin{itemize}
    \item Vector fields along $\rho^{ab}$ which we group according to the homogeneity in $\xi$. We have:
    \begin{equation}
        Y^{(4)}_{ab} = \xi^{(4)}\frac{\partial}{\partial \rho_{ab}}, \qquad Y^{(3)}_l  = \xi^{(3)}_d \frac{\partial}{\partial \rho_{dl}}\,.
    \end{equation}
    Next there are fields which are second order on $\xi$:
    \begin{equation}
        Y^{(2ant)}_{ld} = \epsilon_{abcd}\xi^c \xi_l \frac{\partial}{\partial \rho_{ab}}\,, \qquad
        Y^{(2trace)}_{ld} = \xi^c\xi_l \delta^{ab}_{cd} \frac{\partial}{\partial \rho_{ab}} \,.
    \end{equation}
    Note that these are linearly dependent as $Y^{(2ant)}_{ld} \epsilon^{ldkp} = Y^{(2trace)}{}^{kp}$. The linear in $\xi$ ones are:
    \begin{equation}
    \begin{aligned}
        Y^{(ant)}_d &= \epsilon_{abcd}\xi^c \frac{\partial}{\partial \rho_{ab}} + W_{dnm}{ }^a\epsilon^{nmbc}\frac{\partial}{\partial C^{abc}} \,,\\
        Y^{(trace)}_d &= \xi^c \delta^{ab}_{cd} \frac{\partial}{\partial \rho_{ab}} - W^{adbc}\frac{\partial}{\partial C^{abc}}\,.    
    \end{aligned}
        \end{equation}
    To see that $Y^{(ant)}_d$ belong to the kernel we observe that:
    
    \begin{multline}
    i_{\epsilon_{abcd}\xi^c \frac{\partial}{\partial \rho_{ab}}} \omega_W = \epsilon_{abcd}\xi^c d(W^{abnm}\epsilon_{nmpk}\xi^p\xi^k) = \\ = 4\xi^c W_{cdab}d(\xi^a \xi^b) = 8W_{dcba}d(\xi^a)\xi^c \xi^b\,.
    \end{multline}
    This term cancels with 
    \begin{equation}
        i_{Y^{(ant)}_d}(-2\omega_{C}) = -2(d(\xi^a) W_{dnma} \epsilon^{nmbc}\epsilon_{bcpk}\xi^p \xi^k )\,.
    \end{equation}
    The considerations for $Y^{(trace)}_d$, $Y^{(2ant)}_{ld}$, and $Y^{(2trace)}_{ld}$ are pretty much the same.

\item Vector fields along $\xi^a$. We also group them according to the homogeneity in $\xi$:
    \begin{equation}
        X^{(4)}_a = \xi^{(4)}\frac{\partial}{\partial \xi_{a}}\,, \qquad
        X^{(3)} = \xi^{(3)}_a\frac{\partial}{\partial \xi_{a}}\,.
    \end{equation}
    Second order ones :
    \begin{equation}
    \begin{gathered}
        X^{(2ant)}_d = \epsilon_{abcd}\xi^b \xi^c \frac{\partial}{\partial \xi_{a}} - \xi_d W^{ab}{ }_{pk}\epsilon^{pknm}\frac{\partial}{\partial W^{abnm}} -\\ - 4\xi_d C^a{ }_{pk}\epsilon^{pkbc}\frac{\partial}{\partial C^{abc}} - 4\xi^a C_{pkd} \epsilon^{bcpk} \frac{\partial}{\partial C^{abc}}\,, \\ X^{(2trace)}_d = \xi_b \xi_c \delta^{bc}_{ad}  \frac{\partial}{\partial \xi_{a}} - \xi_d W^{abnm}\frac{\partial}{\partial W^{abnm}} -\\ - 6\xi_d C^{abc}\frac{\partial}{\partial C^{abc}}\,. 
    \end{gathered}
    \end{equation}
    To see that they are in the kernel we observe the following relations:
    \begin{multline}
        i_{\epsilon_{aljd}\xi^l \xi^j \frac{\partial}{\partial \xi_{a}}} \omega_W =  2d(\rho_{ab})W^{abnm}\epsilon_{nmpk}\epsilon^{pljd}\xi_l \xi_j \xi^k = \\ = 2d(\rho_{ab})W^{abnm}\delta^{ljd}_{nmk} \xi_l \xi_j \xi^k  = 4d(\rho_{ab})W^{abnm} \xi_n \xi_m \xi^d \,,
    \end{multline}
    \begin{multline}
        i_{\epsilon_{aljd}\xi^l \xi^j \frac{\partial}{\partial \xi_{a}}} \omega_{C} = \epsilon_{aljd}\xi^l \xi^j d(C^a{ }_{bc}\epsilon^{bcpk}\xi_p \xi_k) + 2 d(\xi_a) C^a{ }_{bc}\epsilon^{bcpk} \epsilon_{pljd}\xi^l \xi^j \xi_k = \\ = 2\delta_{ad}^{pk}d(C^a{ }_{bc})\epsilon^{bcpk} + 2 \delta_{aljd}^{bcpk}\xi^l \xi^j C^a{ }_{bc} d(\xi_p) \xi_k + 2 d(\xi_a)C^a{ }_{bc}\delta^{bck}_{ljd}\xi^l \xi^j \xi_k = \\ =
        4\xi^b \xi^c C^a{ }_{bc} d(\xi_a) \xi_d + 4 \xi_a C^a{ }_{bd} \xi^b \xi^k d(\xi_k) \,.
    \end{multline}
    
    And finally the first order one is given by:
    \begin{equation}
        X^{(trace)} = \xi^a\frac{\partial}{\partial \xi_{a}} - 2 W^{abnm}\frac{\partial}{\partial W^{abnm}} - 3 C^{abc}\frac{\partial}{\partial C^{abc}}\,.
    \end{equation}
    The proof that it is the kernel is based on:
    \begin{multline} \label{xidxi}
        i_{\xi^a\frac{\partial}{\partial \xi_{a}}}\omega = 2d(\rho_{ab})W^{abnm}\epsilon_{nmpk}\xi^p\xi^k -\\- 2(\xi_ad(C^a{ }_{bc}\epsilon^{bcpk}\xi_p \xi_k) + 2d(\xi_a)C^a{ }_{bc}\epsilon^{bcpk} \xi_p \xi_k)\,.
    \end{multline}
    The first term is canceled by $i_{-2 W^{abnm}\frac{\partial}{\partial W^{abnm}}}\omega_W$. Then $\xi_ad(C^a{ }_{bc})\epsilon^{bcpk}\xi_p \xi_k = 0$ due to Bianchi identity.  In terms of $\stackrel{*}{C^{apk}} = C^a{ }_{bc}\epsilon^{bcpk}$ the second term gives:
    \begin{multline}
        \xi_a C^a{ }_{bc}\epsilon^{bcpk}d(\xi_p) \xi_k - \xi_a C^a{ }_{bc}\epsilon^{bcpk}\xi_p d(\xi_k) = \stackrel{*}{C^{apk}}(\xi_ad(\xi_p) \xi_k - \xi_a \xi_p d(\xi_k)) = \\ = \stackrel{*}{C^{apk}} \xi_a d(\xi_p) \xi_k + \stackrel{*}{C^{apk}}\xi_p d(\xi_k) \xi_a  = (\stackrel{*}{C^{apk}} + \stackrel{*}{C^{kap}})(\xi_a d(\xi_p) \xi_k) = \\ = (-\stackrel{*}{C^{pka}})(\xi_a d(\xi_p) \xi_k) = \stackrel{*}{C^{pka}}( d(\xi_p) \xi_k  \xi_a)\,.
    \end{multline}
    Together with the third terms in \eqref{xidxi} they can be taken care of by the $-3 C^{abc}\frac{\partial}{\partial C^{abc}}$.

\item Finally we have two more independent kernel vector fields 
\begin{equation}
    L= \frac{\partial}{\partial \lambda}\,, \qquad
    K = \frac{\partial}{\partial \kappa}\,,
\end{equation}
simply because neither $d\kappa^a$ nor $d\lambda$ enter
\eqref{presymp-main}.
\end{itemize}

Now we consider the symplectic structure $\bar\omega$ on $\bar F$ at a point $p$, where all the coordinates but $e^a_\mu$ are set to zero. At this point $\bar\omega$ reads as:
\begin{equation} \label{symp_struct_at_pointAp}
\begin{gathered}
    \bar{\omega}_p =  (d w_{ab}d \st{2}{W} {}^{abcd}{ }_{,cd} + d\omega_{ab,c} d \st{1}{W}{}^{abcd}{ }_{,d} + d \omega_{ab,cd} d W^{abcd}) - \\ -2(d \epsilon_{a} d \st{2}{C}{}^{acd}{ }_{,cd} + d e_{a,c} d \st{1}{C}{}^{acd}{ }_{,d} + d e_{a,cd} d C^{acd})\,,
\end{gathered}
\end{equation}
in a suitable coordinate system. By inspecting the prolongation of the above kernel vector fields $X,Y,K,L$ at this point one finds 
that their linear span coincides with the kernel. The argument given in the main text then implies that their prolongations generate the kernel distribution of $\bar\omega$.



\section{Metric elimination} \label{MetrEl}

Here we discuss the assumptions under which the metric can be eliminated. As we discuss in the main text the BRST complex of CGR as presented in \cite{Boulanger:2004eh} contain ghost degree $1$ elements  ${\xi}_\mu{ }^\nu$ and the metric.
Thanks to 
\begin{equation}
    \gamma g_{\mu\nu} = -2\lambda g_{\mu\nu} + \xi_\mu{ }^\rho g_{\rho \nu} + {\xi}_\nu{ }^\rho g_{\mu \rho} =-2\lambda g_{\mu \nu} + {\xi}_{(\mu \nu)}\,,
\end{equation}
$g_{\mu\nu}$ and ${\xi}_{(\mu \nu)}$ can be eliminated at least locally.

However, $g_{\mu\nu}$ are not unconstrained coordinates in the sense of definition \bref{contr_man} because the determinant of the metric can not be zero. Moreover, the space is further restricted to the space of symmetric matrices of definite signature. 

It turns our that in the case of Riemannian metric (i.e. Euclidean signature) the metric can be eliminated. To see this we employ Cholesky decomposition which states that every symmetric, positive-definite matrix can be uniquely decomposed as 
\begin{equation}
    g = LL^T\,,
\end{equation}
where L is a lower-triangular matrix with positive elements on the diagonal. The space of such matrices is diffeomorphic to $R^{n(n+1)/2}$ and the concrete diffeomorphism can be obtained by taking the coordinates on $R^{n(n+1)/2}$ to be the entries of a lower-triangular matrix $A$ and taking $L = e^A$. Entries of the matrix $A$ then become the free coordinates parameterizing the space of metrics and form contractible pairs with the symmetric part of $\xi_{\mu\nu}$. 

In the case of pseudo-Euclidean metrics there might be no such decomposition. We will make use of the following theorem \cite{GoluVanl96} Th. 4.1.1, 4.1.2

\begin{thm} \label{thm:metric-decom}
If all the leading principal minors of $A \in R^{n \times n}$ are nonvanishing then there exist a unique lower triangular matrix $L$ with positive diagonal values and a unique diagonal matrix $D = diag(d_1, ..., d_n), d_k= \pm 1$ such that $A = LDL^T$.
\end{thm}
A metric in general coordinates might not satisfy Theorem \ref{thm:metric-decom}. For example take the metric on a plane in $z$ coordinates: $ds^2 = dzd\Bar{z}$.  If we restrict to such metrics that admit such factorization the elimination can still be performed.  This subspace of metrics includes important cases, for example all metrics close to Minkowsky one. 

\section{Presymplectic representation of the standard \\ "gauge" formulation} \label{sec:Kaku-str}
The simple approach one might try to use to find a presymplectic AKSZ formulation of CRR is to consider just the conformal algebra with shifted grading $ g[1]$ as the target space with the canonical Chevalie-Eilenberg differential $q_{CE}$ on it.
The presymplectic potential and structure can be read off the action \eqref{action} presented in \cite{Kaku:1977pa}
\begin{equation}
\begin{gathered}
    \chi_G = (d\rho_{ab} \xi_c \kappa_d) \epsilon^{abcd}  = (\xi_a \kappa_b d\rho_{cd}) \epsilon^{abcd}\,, \\
    \omega_{G} = d\chi = (d\xi_a \kappa_b d\rho_{cd} - \xi_a d\kappa_b d\rho_{cd})\epsilon^{abcd}.
\end{gathered}
\end{equation}
This structure is invariant under $q_{CE}$ and induces the following action:
\begin{equation}
    S = \int(d\omega_{ab}e_c f_d + \omega_{a}{ }^{k}\omega_{kb}e_c f_d + e_a f_be_c f_d)\epsilon^{abcd}.
\end{equation}

This procedure reproduces the action \eqref{action} but for it to be equivalent to the conformal gravity in metric formulation one has to additionally impose the zero torsion condition.

\bibliographystyle{utphys}
\setlength{\itemsep}{0em}
\small

\begin{thebibliography}{10}

\bibitem{Alexandrov:1995kv}
M.~Alexandrov, M.~Kontsevich, A.~Schwartz, and O.~Zaboronsky, ``{T}he
  {G}eometry of the master equation and topological quantum field theory,''
  \href{http://dx.doi.org/10.1142/S0217751X97001031}{{\em Int.J.Mod.Phys.}
  {\bfseries A12} (1997) 1405--1430},
\href{http://arxiv.org/abs/hep-th/9502010}{{\ttfamily hep-th/9502010}}.

\bibitem{Alkalaev:2013hta}
K.~B. Alkalaev and M.~Grigoriev, ``{Frame-like Lagrangians and presymplectic
  AKSZ-type sigma models},''
  \href{http://dx.doi.org/10.1142/S0217751X14501036}{{\em Int. J. Mod. Phys.}
  {\bfseries A29} no.~18, (2014) 1450103},
\href{http://arxiv.org/abs/1312.5296}{{\ttfamily arXiv:1312.5296 [hep-th]}}.

\bibitem{Grigoriev:2016wmk}
M.~Grigoriev, ``{Presymplectic structures and intrinsic Lagrangians},''
\href{http://arxiv.org/abs/1606.07532}{{\ttfamily arXiv:1606.07532 [hep-th]}}.

\bibitem{Kijowski:1979dj}
J.~Kijowski and W.~M. Tulczyjew, {\em {A symplectic framework for field
  theories }}.
\newblock 1979.

\bibitem{Crnkovic:1986ex}
C.~Crnkovic and E.~Witten, ``{C}ovariant {D}escription {O}f {C}anonical
  {F}ormalism {I}n {G}eometrical {T}heories,''.
in Three hundred years of gravitation, S. W. Hawking and W. Israel, eds., pp.
  676-684. Cambridge University Press, Cambridge, 1987.

\bibitem{Khavkine2012}
I.~Khavkine, ``{P}resymplectic current and the inverse problem of the calculus
  of variations,'' \href{http://dx.doi.org/10.1063/1.4828666}{{\em J. Math.
  Phys.} {\bfseries 54,} (Oct., 2012) 111502},
  \href{http://arxiv.org/abs/1210.0802}{{\ttfamily 1210.0802}}.

\bibitem{Sharapov:2016sgx}
A.~A. Sharapov, ``{Variational tricomplex, global symmetries and conservation
  laws of gauge systems},''
\href{http://arxiv.org/abs/1607.01626}{{\ttfamily arXiv:1607.01626 [math-ph]}}.

\bibitem{Grigoriev:2020xec}
M.~Grigoriev and A.~Kotov, ``{Presymplectic AKSZ formulation of Einstein
  gravity},'' \href{http://arxiv.org/abs/2008.11690}{{\ttfamily
  arXiv:2008.11690 [hep-th]}}.

\bibitem{Batalin:1981jr}
I.~Batalin and G.~Vilkovisky, ``{G}auge {A}lgebra and {Q}uantization,''
\href{http://dx.doi.org/10.1016/0370-2693(81)90205-7}{{\em Phys.Lett.}
  {\bfseries B102} (1981) 27--31}.

\bibitem{Batalin:1983wj}
I.~Batalin and G.~Vilkovisky, ``{F}eynman {R}ules {F}or {R}educible {G}auge
  {T}heories,''
\href{http://dx.doi.org/10.1016/0370-2693(83)90645-7}{{\em Phys.Lett.}
  {\bfseries B120} (1983) 166--170}.

\bibitem{Canepa:2020rhu}
G.~Canepa, A.~S. Cattaneo, and M.~Schiavina, ``{General Relativity and the AKSZ
  construction},'' \href{http://arxiv.org/abs/2006.13078}{{\ttfamily
  arXiv:2006.13078 [math-ph]}}.

\bibitem{Sharapov:2021drr}
A.~Sharapov and E.~Skvortsov, ``{Higher spin gravities and presymplectic AKSZ
  models},'' \href{http://dx.doi.org/10.1016/j.nuclphysb.2021.115551}{{\em
  Nucl. Phys. B} {\bfseries 972} (2021) 115551},
  \href{http://arxiv.org/abs/2102.02253}{{\ttfamily arXiv:2102.02253
  [hep-th]}}.

\bibitem{Kaku:1977pa}
M.~Kaku, P.~K. Townsend, and P.~van Nieuwenhuizen, ``{G}auge {T}heory of the
  {C}onformal and {S}uperconformal {G}roup,''
\href{http://dx.doi.org/10.1016/0370-2693(77)90552-4}{{\em Phys. Lett.}
  {\bfseries B69} (1977) 304--308}.

\bibitem{Fradkin:1985am}
E.~S. Fradkin and A.~A. Tseytlin, ``{C}onformal {S}upergravity,''
\href{http://dx.doi.org/10.1016/0370-1573(85)90138-3}{{\em Phys. Rept.}
  {\bfseries 119} (1985) 233--362}.

\bibitem{Maldacena:2011mk}
J.~Maldacena, ``{Einstein Gravity from Conformal Gravity},''
  \href{http://arxiv.org/abs/1105.5632}{{\ttfamily arXiv:1105.5632 [hep-th]}}.

\bibitem{Mannheim:2011ds}
P.~D. Mannheim, ``{Making the Case for Conformal Gravity},''
  \href{http://dx.doi.org/10.1007/s10701-011-9608-6}{{\em Found. Phys.}
  {\bfseries 42} (2012) 388--420},
  \href{http://arxiv.org/abs/1101.2186}{{\ttfamily arXiv:1101.2186 [hep-th]}}.

\bibitem{Rachwal:2022pfe}
L.~Rachwa\l{}, ``{Introduction to Quantization of Conformal Gravity},''
  \href{http://dx.doi.org/10.3390/universe8040225}{{\em Universe} {\bfseries 8}
  no.~4, (2022) 225}, \href{http://arxiv.org/abs/2204.13856}{{\ttfamily
  arXiv:2204.13856 [hep-th]}}.

\bibitem{Edery:2006hg}
A.~Edery, L.~Fabbri, and M.~B. Paranjape, ``{Spontaneous breaking of conformal
  invariance in theories of conformally coupled matter and Weyl gravity},''
  \href{http://dx.doi.org/10.1088/0264-9381/23/22/019}{{\em Class. Quant.
  Grav.} {\bfseries 23} (2006) 6409--6423},
  \href{http://arxiv.org/abs/hep-th/0603131}{{\ttfamily arXiv:hep-th/0603131}}.

\bibitem{Boulanger:2004eh}
N.~Boulanger, ``{A} {W}eyl-covariant tensor calculus,''
  \href{http://dx.doi.org/10.1063/1.1896381}{{\em J. Math. Phys.} {\bfseries
  46} (2005) 053508},
\href{http://arxiv.org/abs/hep-th/0412314}{{\ttfamily arXiv:hep-th/0412314}}.

\bibitem{Boulanger:2007st}
N.~Boulanger, ``{General solutions of the Wess-Zumino consistency condition for
  the Weyl anomalies},''
  \href{http://dx.doi.org/10.1088/1126-6708/2007/07/069}{{\em JHEP} {\bfseries
  07} (2007) 069},
\href{http://arxiv.org/abs/0704.2472}{{\ttfamily arXiv:0704.2472 [hep-th]}}.

\bibitem{Boulanger:2007ab}
N.~Boulanger, ``{Algebraic Classification of Weyl Anomalies in Arbitrary
  Dimensions},'' \href{http://dx.doi.org/10.1103/PhysRevLett.98.261302}{{\em
  Phys. Rev. Lett.} {\bfseries 98} (2007) 261302},
\href{http://arxiv.org/abs/0706.0340}{{\ttfamily arXiv:0706.0340 [hep-th]}}.

\bibitem{Joung:2021bhf}
E.~Joung, M.-g. Kim, and Y.~Kim, ``{Unfolding conformal geometry},''
  \href{http://dx.doi.org/10.1007/JHEP12(2021)092}{{\em JHEP} {\bfseries 12}
  (2021) 092}, \href{http://arxiv.org/abs/2108.05535}{{\ttfamily
  arXiv:2108.05535 [hep-th]}}.

\bibitem{Preitschopf:1998ei}
C.~Preitschopf and M.~A. Vasiliev, ``{C}onformal field theory in conformal
  space,'' \href{http://dx.doi.org/10.1016/S0550-3213(99)00087-5}{{\em
  Nucl.Phys.} {\bfseries B549} (1999) 450--480},
\href{http://arxiv.org/abs/hep-th/9812113}{{\ttfamily arXiv:hep-th/9812113}}.

\bibitem{Grigoriev:2019ojp}
M.~Grigoriev and A.~Kotov, ``{Gauge PDE and AKSZ-type Sigma Models},''
  \href{http://dx.doi.org/10.1002/1521-3978(200209)50:8/9<825::AID-PROP825>3.0.CO;2-V}{{\em
  Fortsch. Phys.} (2019) }, \href{http://arxiv.org/abs/1903.02820}{{\ttfamily
  arXiv:1903.02820 [hep-th]}}.

\bibitem{Barnich:2010sw}
G.~Barnich and M.~Grigoriev, ``{F}irst order parent formulation for generic
  gauge field theories,'' \href{http://dx.doi.org/10.1007/JHEP01(2011)122}{{\em
  JHEP} {\bfseries 01} (2011) 122},
\href{http://arxiv.org/abs/1009.0190}{{\ttfamily arXiv:1009.0190 [hep-th]}}.

\bibitem{Grigoriev:2010ic}
M.~Grigoriev, ``{P}arent formulation at the {L}agrangian level,''
  \href{http://dx.doi.org/10.1007/JHEP07(2011)061}{{\em JHEP} {\bfseries 07}
  (2011) 061},
\href{http://arxiv.org/abs/1012.1903}{{\ttfamily arXiv:1012.1903 [hep-th]}}.

\bibitem{Kotov:2007nr}
A.~Kotov and T.~Strobl, ``{Characteristic classes associated to Q-bundles},''
  \href{http://dx.doi.org/10.1142/S0219887815500061}{{\em Int. J. Geom. Meth.
  Mod. Phys.} {\bfseries 12} no.~01, (2014) 1550006},
\href{http://arxiv.org/abs/0711.4106}{{\ttfamily arXiv:0711.4106 [math.DG]}}.

\bibitem{Grigoriev:1999qz}
M.~A. Grigoriev and P.~H. Damgaard, ``{S}uperfield {BRST} charge and the master
  action,'' \href{http://dx.doi.org/10.1016/S0370-2693(00)00050-2}{{\em Phys.
  Lett.} {\bfseries B474} (2000) 323--330},
\href{http://arxiv.org/abs/hep-th/9911092}{{\ttfamily arXiv:hep-th/9911092
  [hep-th]}}.

\bibitem{Barnich:1995db}
G.~Barnich, F.~Brandt, and M.~Henneaux, ``{L}ocal {BRST} cohomology in the
  antifield formalism. {I}. {G}eneral theorems,'' {\em Commun. Math. Phys.}
  {\bfseries 174} (1995) 57--92,
\href{http://arxiv.org/abs/hep-th/9405109}{{\ttfamily hep-th/9405109}}.

\bibitem{Brandt:1996mh}
F.~Brandt, ``{L}ocal {BRST} {C}ohomology and {C}ovariance,'' {\em Commun. Math.
  Phys.} {\bfseries 190} (1997) 459--489,
\href{http://arxiv.org/abs/hep-th/9604025}{{\ttfamily hep-th/9604025}}.

\bibitem{Grigoriev:2012xg}
M.~Grigoriev, ``{P}arent formulations, frame-like {L}agrangians, and
  generalized auxiliary fields,''
  \href{http://dx.doi.org/10.1007/JHEP12(2012)048}{{\em JHEP} {\bfseries 1212}
  (2012) 048},
\href{http://arxiv.org/abs/1204.1793}{{\ttfamily arXiv:1204.1793 [hep-th]}}.

\bibitem{Trujillo:2013saa}
J.~T. Trujillo, {\em {Weyl Gravity as a Gauge Theory}}.
\newblock PhD thesis, Utah State U., 2013.

\bibitem{Barnich:1995ap}
G.~Barnich, F.~Brandt, and M.~Henneaux, ``{L}ocal {BRST} cohomology in
  {E}instein {Y}ang-{M}ills theory,'' {\em Nucl. Phys.} {\bfseries B455} (1995)
  357--408,
\href{http://arxiv.org/abs/hep-th/9505173}{{\ttfamily hep-th/9505173}}.

\bibitem{Grigoriev:2020lzu}
M.~Grigoriev, K.~Mkrtchyan, and E.~Skvortsov, ``{Matter-free higher spin
  gravities in 3D: Partially-massless fields and general structure},''
  \href{http://dx.doi.org/10.1103/PhysRevD.102.066003}{{\em Phys. Rev. D}
  {\bfseries 102} no.~6, (2020) 066003},
  \href{http://arxiv.org/abs/2005.05931}{{\ttfamily arXiv:2005.05931
  [hep-th]}}.

\bibitem{Stora:1983ct}
R.~Stora, ``Algebraic structure and topological origin of anomalies,''. Seminar
  given at Cargese Summer Inst.: Progress in Gauge Field Theory, Cargese,
  France, Sep 1-15, 1983.

\bibitem{Manes:1985df}
J.~Manes, R.~Stora, and B.~Zumino, ``{Algebraic Study of Chiral Anomalies},''
  \href{http://dx.doi.org/10.1007/BF01208825}{{\em Commun. Math. Phys.}
  {\bfseries 102} (1985) 157}.

\bibitem{Vasiliev:1988xc}
M.~A. Vasiliev, ``{Equations of Motion of Interacting Massless Fields of All
  Spins as a Free Differential Algebra},''
\href{http://dx.doi.org/10.1016/0370-2693(88)91179-3}{{\em Phys. Lett.}
  {\bfseries B209} (1988) 491--497}.

\bibitem{Barnich:2004cr}
G.~Barnich, M.~Grigoriev, A.~Semikhatov, and I.~Tipunin, ``{P}arent field
  theory and unfolding in {BRST} first-quantized terms,''
  \href{http://dx.doi.org/10.1007/s00220-005-1408-4}{{\em Commun.Math.Phys.}
  {\bfseries 260} (2005) 147--181},
\href{http://arxiv.org/abs/hep-th/0406192}{{\ttfamily arXiv:hep-th/0406192
  [hep-th]}}.

\bibitem{Henneaux:1990ua}
M.~Henneaux, ``{Elimination of the Auxiliary Fields in the Antifield
  Formalism},''
\href{http://dx.doi.org/10.1016/0370-2693(90)91739-X}{{\em Phys. Lett.}
  {\bfseries B238} (1990) 299}.

\bibitem{Barnich:2000zw}
G.~Barnich, F.~Brandt, and M.~Henneaux, ``{L}ocal {BRST} cohomology in gauge
  theories,'' \href{http://dx.doi.org/10.1016/S0370-1573(00)00049-1}{{\em
  Phys.Rept.} {\bfseries 338} (2000) 439--569},
\href{http://arxiv.org/abs/hep-th/0002245}{{\ttfamily hep-th/0002245}}.

\bibitem{Brandt:2001tg}
F.~Brandt, ``{J}et coordinates for local {BRST} cohomology,''
  \href{http://dx.doi.org/10.1023/A:1010917617033}{{\em Lett. Math. Phys.}
  {\bfseries 55} (2001) 149--159},
\href{http://arxiv.org/abs/math-ph/0103006}{{\ttfamily arXiv:math-ph/0103006}}.

\bibitem{Berkovits:2004jj}
N.~Berkovits and E.~Witten, ``{Conformal supergravity in twistor-string
  theory},'' \href{http://dx.doi.org/10.1088/1126-6708/2004/08/009}{{\em JHEP}
  {\bfseries 08} (2004) 009},
  \href{http://arxiv.org/abs/hep-th/0406051}{{\ttfamily arXiv:hep-th/0406051}}.

\bibitem{Adamo:2016ple}
T.~Adamo, P.~Hähnel, and T.~McLoughlin, ``{Conformal higher spin scattering
  amplitudes from twistor space},''
  \href{http://dx.doi.org/10.1007/JHEP04(2017)021}{{\em JHEP} {\bfseries 04}
  (2017) 021},
\href{http://arxiv.org/abs/1611.06200}{{\ttfamily arXiv:1611.06200 [hep-th]}}.

\bibitem{Hahnel:2016ihf}
P.~H\"ahnel and T.~McLoughlin, ``{Conformal higher spin theory and twistor
  space actions},'' \href{http://dx.doi.org/10.1088/1751-8121/aa9108}{{\em J.
  Phys. A} {\bfseries 50} no.~48, (2017) 485401},
  \href{http://arxiv.org/abs/1604.08209}{{\ttfamily arXiv:1604.08209
  [hep-th]}}.

\bibitem{Krasnov:2021nsq}
K.~Krasnov, E.~Skvortsov, and T.~Tran, ``{Actions for self-dual Higher Spin
  Gravities},'' \href{http://dx.doi.org/10.1007/JHEP08(2021)076}{{\em JHEP}
  {\bfseries 08} (2021) 076}, \href{http://arxiv.org/abs/2105.12782}{{\ttfamily
  arXiv:2105.12782 [hep-th]}}.

\bibitem{Vasiliev:2005zu}
M.~A. Vasiliev, ``{A}ctions, charges and off-shell fields in the unfolded
  dynamics approach,'' {\em Int. J. Geom. Meth. Mod. Phys.} {\bfseries 3}
  (2006) 37--80,
\href{http://arxiv.org/abs/hep-th/0504090}{{\ttfamily hep-th/0504090}}.

\bibitem{Batalin:1998pz}
I.~A. Batalin, K.~Bering, and P.~H. Damgaard, ``{S}uperfield formulation of the
  phase space path integral,''
  \href{http://dx.doi.org/10.1016/S0370-2693(98)01537-8}{{\em Phys. Lett.}
  {\bfseries B446} (1999) 175--178},
\href{http://arxiv.org/abs/hep-th/9810235}{{\ttfamily arXiv:hep-th/9810235}}.

\bibitem{Batalin:1997ks}
I.~A. Batalin, K.~Bering, and P.~H. Damgaard, ``{S}uperfield quantization,''
  \href{http://dx.doi.org/10.1016/S0550-3213(97)00806-7}{{\em Nucl. Phys.}
  {\bfseries B515} (1998) 455--487},
\href{http://arxiv.org/abs/hep-th/9708140}{{\ttfamily arXiv:hep-th/9708140}}.

\bibitem{Grigoriev:2021wgw}
M.~Grigoriev and V.~Gritzaenko, ``{Presymplectic structures and intrinsic
  Lagrangians for massive fields},''
  \href{http://dx.doi.org/10.1016/j.nuclphysb.2022.115686}{{\em Nuclear Physics
  B} {\bfseries 975} (9, 2021) 115686},
  \href{http://arxiv.org/abs/2109.05596}{{\ttfamily arXiv:2109.05596
  [hep-th]}}.

\bibitem{Zuckerman:1989cx}
G.~J. Zuckerman, ``{A}ction principles and global geometry,''
{\em Conf. Proc.} {\bfseries C8607214} (1986) 259--284.

\bibitem{Anderson1991}
I.~Anderson, ``{I}ntroduction to the variational bicomplex,'' in {\em
  Mathematical Aspects of Classical Field Theory}, M.~Gotay, J.~Marsden, and
  V.~Moncrief, eds., vol.~132 of {\em Contemporary Mathematics}, pp.~51--73.
\newblock Amer. Math. Soc., 1992.

\bibitem{Grigoriev:2016bzl}
M.~Grigoriev and A.~A. Tseytlin, ``{On conformal higher spins in curved
  background},'' \href{http://dx.doi.org/10.1088/1751-8121/aa5c5f}{{\em J.
  Phys.} {\bfseries A50} no.~12, (2017) 125401},
\href{http://arxiv.org/abs/1609.09381}{{\ttfamily arXiv:1609.09381 [hep-th]}}.

\bibitem{Lopatin:1987hz}
V.~E. Lopatin and M.~A. Vasiliev, ``{Free Massless Bosonic Fields of Arbitrary
  Spin in $d$-dimensional De Sitter Space},''
\href{http://dx.doi.org/10.1142/S0217732388000313}{{\em Mod. Phys. Lett.}
  {\bfseries A3} (1988) 257}.

\bibitem{Chalmers:1996rq}
G.~Chalmers and W.~Siegel, ``{The Selfdual sector of QCD amplitudes},''
  \href{http://dx.doi.org/10.1103/PhysRevD.54.7628}{{\em Phys. Rev. D}
  {\bfseries 54} (1996) 7628--7633},
  \href{http://arxiv.org/abs/hep-th/9606061}{{\ttfamily arXiv:hep-th/9606061}}.

\bibitem{Curry:2014yoa}
S.~Curry and A.~R. Gover, ``{An introduction to conformal geometry and tractor
  calculus, with a view to applications in general relativity},''
  \href{http://arxiv.org/abs/1412.7559}{{\ttfamily arXiv:1412.7559 [math.DG]}}.

\bibitem{Fradkin:1989md}
E.~S. Fradkin and V.~Y. Linetsky, ``{Cubic Interaction in Conformal Theory of
  Integer Higher Spin Fields in Four-dimensional Space-time},''
  \href{http://dx.doi.org/10.1016/0370-2693(89)90120-2}{{\em Phys. Lett. B}
  {\bfseries 231} (1989) 97--106}.

\bibitem{Beccaria:2014jxa}
M.~Beccaria, X.~Bekaert, and A.~A. Tseytlin, ``{P}artition function of free
  conformal higher spin theory,''
  \href{http://dx.doi.org/10.1007/JHEP08(2014)113}{{\em JHEP} {\bfseries 1408}
  (June, 2014) 113},
\href{http://arxiv.org/abs/1406.3542}{{\ttfamily 1406.3542}}.

\bibitem{Bekaert:2013zya}
X.~Bekaert and M.~Grigoriev, ``{H}igher order singletons, partially massless
  fields and their boundary values in the ambient approach,''
  \href{http://dx.doi.org/10.1016/j.nuclphysb.2013.08.015}{{\em Nucl.Phys.}
  {\bfseries B876} (2013) 667--714},
\href{http://arxiv.org/abs/1305.0162}{{\ttfamily arXiv:1305.0162 [hep-th]}}.

\bibitem{Kuzenko:2019ill}
S.~M. Kuzenko and M.~Ponds, ``{Conformal geometry and (super)conformal
  higher-spin gauge theories},''
  \href{http://dx.doi.org/10.1007/JHEP05(2019)113}{{\em JHEP} {\bfseries 05}
  (2019) 113},
\href{http://arxiv.org/abs/1902.08010}{{\ttfamily arXiv:1902.08010 [hep-th]}}.

\bibitem{Segal:2002gd}
A.~Y. Segal, ``{C}onformal higher spin theory,'' {\em Nucl. Phys.} {\bfseries
  B664} (2003) 59--130,
\href{http://arxiv.org/abs/hep-th/0207212}{{\ttfamily hep-th/0207212}}.

\bibitem{Tseytlin:2002gz}
A.~A. Tseytlin, ``{O}n limits of superstring in {A}d{S}(5) x {S}**5,''
  \href{http://dx.doi.org/10.1023/A:1020646014240}{{\em Theor. Math. Phys.}
  {\bfseries 133} (2002) 1376--1389},
\href{http://arxiv.org/abs/hep-th/0201112}{{\ttfamily arXiv:hep-th/0201112}}.

\bibitem{Bekaert:2010ky}
X.~Bekaert, E.~Joung, and J.~Mourad, ``{Effective action in a higher-spin
  background},'' \href{http://dx.doi.org/10.1007/JHEP02(2011)048}{{\em JHEP}
  {\bfseries 02} (2011) 048},
\href{http://arxiv.org/abs/1012.2103}{{\ttfamily arXiv:1012.2103 [hep-th]}}.

\bibitem{Bonezzi:2017mwr}
R.~Bonezzi, ``{Induced Action for Conformal Higher Spins from Worldline Path
  Integrals},'' \href{http://dx.doi.org/10.3390/universe3030064}{{\em Universe}
  {\bfseries 3} no.~3, (2017) 64},
\href{http://arxiv.org/abs/1709.00850}{{\ttfamily arXiv:1709.00850 [hep-th]}}.

\bibitem{Cap:2002aj}
A.~\v{C}ap and A.~R. Gover, ``{S}tandard {T}ractors and the {C}onformal
  {A}mbient {M}etric {C}onstruction,'' {\em Annals Global Anal. Geom.}
  {\bfseries 24} (2003) 231--259,
  \href{http://arxiv.org/abs/math/0207016}{{\ttfamily arXiv:math/0207016}}.
\url{http://dx.doi.org/10.2140/pjm.2006.226.309}.

\bibitem{Krasil'shchik:2010ij}
J.~Krasil'shchik and A.~Verbovetsky, ``{G}eometry of jet spaces and integrable
  systems,'' \href{http://dx.doi.org/10.1016/j.geomphys.2010.10.012}{{\em J.
  Geom. Phys.} {\bfseries 61} (2011) 1633--1674},
\href{http://arxiv.org/abs/1002.0077}{{\ttfamily arXiv:1002.0077 [math.DG]}}.

\bibitem{GoluVanl96}
G.~H. Golub and C.~F. Van~Loan, {\em Matrix Computations}.
\newblock The Johns Hopkins University Press, third~ed., 1996.

\end{thebibliography}

\providecommand{\href}[2]{#2}\begingroup\raggedright\endgroup

\end{document}